   \pdfoutput=1
   \documentclass[fleqn,usenatbib,usedcolumn]{mnras}
   \usepackage[british]{babel}             
\usepackage{amsmath}	
   \usepackage{txfonts}                    
   \let\la=\lesssim 
   \usepackage[T1]{fontenc}                
   \usepackage{graphicx}                   
   \hypersetup{pdfauthor={Maximilian N. G{\"u}nther, Didier Queloz, Brice-Olivier Demory, Francois Bouchy},
               pdftitle={A New Yield Simulator for Transiting Planets and False Positives: Application to the Next Generation Transit Survey},
               pdfkeywords={planets and satellites: detection, eclipses, occultations, surveys, (stars:) binaries: eclipsing, methods: numerical},
               bookmarksnumbered=true}
\usepackage{booktabs}
\usepackage{fancyvrb}

\usepackage{siunitx}
\usepackage[T1]{fontenc}
\usepackage{ae,aecompl}

\title[A New Yield Simulator for Transit Surveys]{A New Yield Simulator for Transiting Planets and False Positives: Application to the Next Generation Transit Survey}

\author[M. N. G{\"u}nther]{Maximilian N. G{\"u}nther$^{1}$\thanks{E-mail: \href{mg719@cam.ac.uk}{mg719@cam.ac.uk}}, Didier Queloz$^{1}$, Brice-Olivier Demory$^{1}$, Francois Bouchy$^{2}$
\\
$^{1}$Astrophysics Group, Cavendish Laboratory, J.J. Thomson Avenue, Cambridge CB3 0HE, UK\\
$^{2}$Observatoire de Gen{\`e}ve, Universit{\'e} de Gen{\`e}ve, 51 Ch. des Maillettes, 1290 Sauverny, Switzerland
}

\date{Last updated -; in original form -}

\pubyear{2016}

\begin{document}
\label{firstpage}
\pagerange{\pageref{firstpage}--\pageref{lastpage}}
\maketitle

\begin{abstract}

We present a yield simulator to predict the number and characteristics of planets, false positives and false alarms in transit surveys.
The simulator is based on a galactic model and the planet occurrence rates measured by the Kepler mission. It takes into account the observation window function and measured noise levels of the investigated survey. Additionally, it includes vetting criteria to identify false positives.
We apply this simulator to the Next Generation Transit Survey (NGTS), a wide-field survey designed to detect transiting Neptune-sized exoplanets. 
We find that red noise is the main limitation of NGTS up to $14$th magnitude, and that its obtained level determines the expected yield.
Assuming a red noise level of $\SI{1}{mmag}$, the simulation predicts the following for a four-year survey: $4\pm3$ Super-Earths, $19\pm5$ Small Neptunes, $16\pm4$ Large Neptunes, $55\pm8$ Saturn-sized planets and $150\pm10$ Jupiter-sized planets, along with $4688\pm45$ eclipsing binaries and $843\pm75$ background eclipsing binaries.
We characterize the properties of these objects to enhance the early identification of false positives and discuss follow-up strategies for transiting candidates.
\end{abstract}

\begin{keywords}
planets and satellites: detection, eclipses, occultations, surveys, (stars:) binaries: eclipsing, methods: numerical
\end{keywords}


\section{Introduction}
Exoplanets transiting their host star give insight into their formation, bulk composition and atmospheric properties. Dedicated wide-field transit surveys, both from the ground (e.g. HAT \citealt{Bakos2002} and WASP \citealt{Pollacco2006}) and from space (e.g. CoRoT \citealt{Baglin2002} and Kepler \citealt{Borucki2010}), have discovered ${\sim}2700$ exoplanets\footnote{\url{http://exoplanetarchive.ipac.caltech.edu/} (17 Aug 2016)}.

Transit-like shape variability in the lightcurve may not only be caused by planets. False alarms introduced by correlated noise may cause a time-periodicity with a pattern similar to a transit shape. In addition, false positive transit events related to an eclipsing astrophysical object can cause a transit signal of small amplitude that may be interpreted as a planetary transit \citep[see e.g.][]{Cameron2012}.
Eclipsing binaries (EBs) can be very expensive in telescope time to follow up.
First, low-mass companions such as Brown dwarfs and very low mass stars can be of similar size as gas giant planets. Distinguishing them from planets necessitates radial velocity follow-up to measure the mass, and may be aided by measuring ellipsoidal effects or obtaining color information during transit and eclipse.
Second, EBs with grazing events lead to transit depths mimicking a planet-sized object even if the secondary is significantly larger.
Another class of false positives are background eclipsing binaries (BEBs), which are faint and distant EBs that are aligned along the line of sight behind a bright target star and hence diluted. The dilution reduces the apparent transit depth onto a planet-like scale, making BEBs one of the most difficult false positives to rule out.
Similar to this are triple and higher-order star systems with one or more pairs of stars eclipsing, referred to as hierarchical EBs.
In wide-field transit surveys, false positives can be up to two orders of magnitude more prevalent than planets \citep[see e.g.][]{Almenara2009,Hartman2011}.

Estimating the yield of a transit experiment provides a way to assess the false positive to planet ratio in detail.
\cite{Brown2003} raised awareness of the contamination impact by false positives in upcoming surveys, but most yield simulations have focused on the number of planets only.
As one of the first, \cite{Brown2008} applied false positive models to predict the yield of the TESS mission. Recently, \cite{Sullivan2015} estimated the planet yield, false positive contamination rates and the success of ad-hoc vetting methods for the TESS mission. 
In addition to enabling insight into future surveys, yield simulations can be used to evaluate how well current instruments achieve their possibilities, as well as how current observing strategies may be optimized.

Here, we develop a yield simulator with the goal of estimating the planet merit and the impact of false signals applicable to any upcoming transit survey. The simulations specifically take into account red noise, false alarms and false positives. 
In order to asses the impact of various observing strategies, the simulation takes as input: target list, telescope parameters, bandpass, field of view, cadence, noise models and detection criteria. To mimic the vetting processes for false positives we implement methods examining the transit parameters (depth, shape and duration), secondary eclipses, centroid movement, and the feasibility of planet follow-up and characterization.

We apply our simulator to estimate the yield of planets and false positives for the recent Next Generation Transit Survey (NGTS) \citep[Wheatley et al., in prep.,][]{Wheatley2013,Chazelas2012}.
Previous ground-based facilities have limited photometric precision, for example $\SIrange{10}{50}{mmag}$ for HAT \citep{Bakos2002} and $\SIrange{3}{30}{mmag}$ for WASP \citep{Pollacco2006}, and are hence more prone to detect Hot Jupiters.
NGTS is designed to be the first ground-based exoplanet survey to reach sub-mmag photometry. It aims at detecting transiting Neptunes with short orbital periods around small stars.
The survey had its first light in early 2015 in Paranal, Chile\footnote{\url{http://www.eso.org/public/news/eso1502/} (17 Aug 2016)}, and started its full science operation in early 2016. 
The facility consists of twelve independent $\SI{20}{cm}$ telescopes with a $\SI{7.4}{sq~deg}$ each, the total field of view adds up to $\SI{88.8}{sq~deg}$, similar to Kepler. NGTS covers a new field of this size every few months, allowing it to survey many bright stars for short orbit planets. 
The sensitivity is optimized between $\SIrange{500}{900}{nm}$ to maximize observation efficiency of K and early-M stars. 

We organize this paper in two major parts. Part one describes the computational layout and the mechanisms of the simulations, which are adoptable to any transit survey. Part two applies the simulations to the example of NGTS.
We describe our simulations and the set of priors in section~\ref{s:Layout of the simulations}. In section~\ref{s:Verifying the simulation on the example of Kepler} we describe the validation process of our code using the results from Kepler. In section~\ref{s:Estimating the yield of NGTS} we examine the case of NGTS, the effects of different red noise levels and detection criteria, and the expected planets and false positives. 
We estimate the feasibility to identify false positives with NGTS' photometric data alone, and provide an outlook into the necessary follow-up facilities for planet candidates. Finally, we discuss our findings and conclude this work in sections~\ref{s:Discussion} and \ref{s:Conclusion}.


\section{Layout of the simulations}
\label{s:Layout of the simulations}
\begin{figure*}
 \includegraphics[width=2.1\columnwidth]{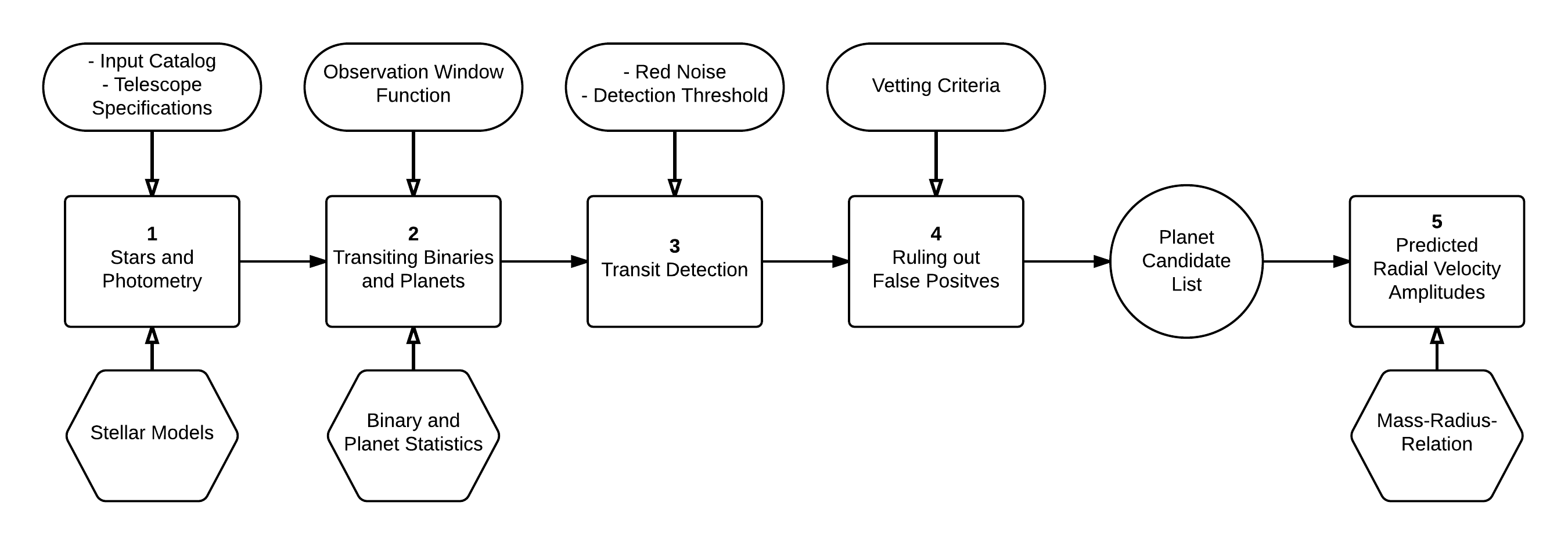}
 \caption{The simulation can be structured in five main modules, containing user input and settings (ovals), observed parameters and models (hexagons), steps of calculations (rectangles), and output files (circle). All user input is given in a parameter file when starting the simulation and is displayed here in respect to where it contributes to the simulation. The numbering refers to sections~\ref{ss:Stars and photometry}-\ref{ss:Doppler follow-up}.}
 \label{fig:flowchart}
\end{figure*}

The overall sketch of the simulation layout as described below is shown in Fig.\ref{fig:flowchart}. 
The simulation input contains a list of stars in the field of view as well as the instrument specifications. An example input file can be found in appendix~\ref{app:Input file for the yield simulation on the example of NGTS}.
From this we calculate the collected flux for each object (section~\ref{ss:Stars and photometry}).
We first randomly assign host stars with planets, and compute the signals of transiting planets as well as eclipsing binaries (depth, duration, shape, visibility, and possible dilution; section~\ref{ss:Transiting planets and binaries}). 
Second, we calculate the total noise for each observation and compute which systems would be detectable (section~\ref{ss:Transit detection}). Third, we rule out false positives if detectable from photometric data (section~\ref{ss:Ruling out false positives}). Finally, we assess the feasibility to follow-up planetary signals with radial velocity instruments (section~\ref{ss:Doppler follow-up}).

\subsection{Stars and photometry}
\label{ss:Stars and photometry}
The simulation considers an input catalog with information about the multiplicity, radius, mass, effective temperature, and magnitude of all stars in the field of view.
The input catalog is built using the TRILEGAL galaxy model \citep{Girardi2005}, up to $V=23$.
We keep the preset adjustments of TRILEGAL referring to the standard Milky Way model and simulate binaries with a fraction of $33\%$ \citep{Raghavan2010}. While this value was estimated for solar-type stars, it is also consistent with predictions for low-mass stars given the dispersion reported in the literature \citep[see e.g.][and references therein]{Duchene2013}. 
The binary mass ratio $q$ of the secondary and primary mass, $q = M_\mathrm{s}/M_\mathrm{p}$, is drawn uniformly between $0.08$ and $1$. To include higher-order multiples we randomly select single stars and assign them to be higher-order multiples.
This way, at the end the input catalog consists of $56\%$ single stars, $33\%$ binaries and $11\%$ higher order multiples \citep{Raghavan2010}. 

From this input catalog we identify a target list of stars according to magnitude and spectral type as listed in Table~\ref{tab:NGTS settings}. 
We assume all input catalog stars are randomly distributed across the field of view.
Then we compute the photometric flux using the effective temperatures and V-band magnitudes of the input catalog and the transmission function of the telescope. 
The stellar parameters are converted into photometric flux using spectrophotometric reference stars from \cite{Pickles1998}.
The zero-point of the V-band is defined by models of Vega from the Kurucz atlas \citep{Kurucz1993}. The Johnson V-band model is adopted from \cite{Buser1978}.
Any other stars lying in the photometric aperture of the target star are considered as background stars for the particular target.

\subsection{Transiting binaries and planets}
\label{ss:Transiting planets and binaries}

\subsubsection{Binaries}
\label{sss:Binaries}
For all input catalog stars identified as binaries, we draw orbital periods $P$ in days from a log-normal distribution with mean of $4.8$ and a standard deviation of $2.3$ \citep{Duquennoy1991}. The eccentricities follow a uniform distribution with a maximum eccentricity given by $e_\mathrm{max} = 0.4 \cdot \log P - 0.2$ \citep[approximated from][]{Raghavan2010}. 
Only detached eclipsing binaries are considered using the Roche limits as criteria. 
Contact eclipsing binaries are evident from the lightcurves and can be readily ruled out.
We do not consider eclipses within triple or higher-order hierarchical eclipsing binaries.

We define orbit and transit parameters following \cite{Winn2011}.
In a binary system, if star 1 is fully transited by star 2, we compute the transit depth as
\begin{align}
\label{eq:binary_transit_full}
\delta_1 &= \frac{A}{A_1} \cdot \frac{F_1}{F_1 + F_2},
\end{align}
where $A = \min{\left(A_1,A_2\right)}$. In the case of a grazing eclipses we replace $A$ with the overlapping area of two circles with radii $R_{1,2}$ and midpoint distance $y$.

\subsubsection{Planets}
\label{sss:Planets}
The planet occurrence rates are based on the results of the Kepler mission as by \cite{Fressin2013} for FGK stars and by \cite{Dressing2015} for small planets around M dwarfs. While recent results made progress on long period and small planets \citep{Burke2015}, in the planet regime targeted by all-sky surveys \cite{Fressin2013} provides to date still the most complete study.
The occurrence rates denote the average number of planets per star binned by planet radius and orbital period. 
In \cite{Fressin2013} they are discretely sampled in radius and period:
Earths ($<1.25~\mathrm{R}_{\oplus}$), Super-Earths ($<2~\mathrm{R}_{\oplus}$), Small Neptunes ($<4~\mathrm{R}_{\oplus}$), Large Neptunes ($<6~\mathrm{R}_{\oplus}$), and giant planets ($<22~\mathrm{R}_{\oplus}$), as well as ten logarithmically spaced period ranges.
We randomly assign a value for period and radius within each discrete interval. The period is drawn within each interval from a logarithmic distribution. We draw the radius within each interval from a uniform distribution, except for the last interval ($>6~\mathrm{R}_{\oplus}$), in which we draw from a logarithmic distribution for consistency with empirical findings \citep[see e.g.][and Exoplanetarchive\footnote{\url{http://exoplanetarchive.ipac.caltech.edu/} (17 Aug 2016)}]{Grether2006}.

We assign planets to all stars, single or binary, in the input catalog according to the occurrence rate (which may be greater than $1$), except when binary systems have orbital periods shorter than $5$~days. Although these short-orbit binary systems are frequent \citep{Slawson2011} and theoretically can host circumbinary planets, these planets are likely undetectable \citep{Munoz2015}.
These constrains lead to a smaller number of assigned planets around close-in binaries than for wide binaries or single stars. Note that there is no evidence that wide binaries affect the occurrence rate of short orbit planets \cite{Deacon2015}.
We do not consider planets around the $11\%$ of higher order multiples in our target list, as in most cases their transit signals will be diluted too much by the other stars in the system to be detectable.
In cases where there is more than one planet assigned to the same star, the orbital parameters of planets are drawn completely independently. This avoids influencing the yield by setting criteria for multiplanetary systems while leading to the same statistical average over all stars of the field.
For all remaining planets in binary systems we compute stability criteria following \cite{Holman1999} and reject planets in unstable orbits ($<1\%$ of all planets).

We investigate the impact of eccentric planetary orbits by assigning various mean eccentricities between $0$ and $0.5$ and find a consistent planet yield as with circular orbits. We therefore employ circular orbits for all planets considering the short orbit sensitivity of all-sky transit surveys. 

We define orbit and transit parameters following \cite{Winn2011}.
For grazing geometry the overlapping area of the two objects is computed instead. If the host system is part of a binary system, we account for dilution by the other star.

\subsubsection{Observation Window Function}
\label{sss:Transit Visibility}
For each transiting planet we estimate the number of transits that can be detected.
The time of the first transit is set randomly between $0$ and the orbital period $P$.
To compute the observation window we calculate the average visibility duration per night of each field.
We implement average weather information at the telescope location and reject a certain fraction of nights to simulate bad conditions.

\subsubsection{Dilution}
Background stars in the aperture of a target star affect its extracted photometry. 
First, they decrease the transit depth of a planet orbiting the target star and might make the signal undetectable. 
Second, if the target star or any background star is an eclipsing binary, its eclipse depth will be decreased and it may appear planet-like. 
The dilution $D$ for a certain source is the ratio of its stellar flux $F_0$ and the total flux in the aperture $\sum_i F_i$. We evaluate $D$ for each system, with $F_0$ being either the flux of a single star or the whole multiple system in which the transit/eclipse occurs. The theoretical transit depth $\delta_0$ from sections \ref{sss:Binaries}-\ref{sss:Planets} is then reduced to the measured transit depth
\begin{align}
\delta = \delta_0 \cdot \frac{F_0}{\sum_i F_i}.
\end{align}

\subsection{Transit detection}
\label{ss:Transit detection}

The total noise of the extracted photometric flux can be described as a composition of uncorrelated and correlated noise, referred to as white and red noise \citep{Pont2006}.
White noise scales with exposure time, aperture, and flux. We calculate it as the sum of individual white noise sources,
\begin{align}
\label{eq:white_noise}
\sigma_\mathrm{white}^2 = t_\mathrm{exp} N_\star + n_\mathrm{pix} \left( t_\mathrm{exp} N_\mathrm{sky} + t_\mathrm{exp} N_\mathrm{dark} + N_\mathrm{read}^2 \right) + \sigma_\mathrm{scint}^2,
\end{align}
where $n_\mathrm{pix}$ is the number of pixels in the aperture, and $t_\mathrm{exp}$ the exposure time. $N_\star$ is the photon count received from a given source and $N_\mathrm{sky}$ the sky background. $N_\mathrm{dark}$ and $N_\mathrm{read}$ are the counts contributed by dark and readout noise. $\sigma_\mathrm{scint}$ describes the scintillation noise, evaluated following \cite{Dravins1998}.

White noise averages out by the square root of the number of exposures per transit, $N_\mathrm{exp}$.
The total noise in one transit is given by the squared quadratic sum of the binned white noise and the red noise. 
Red noise is composed of various sources that are not entirely known, such as weather patterns (if ground-based), correlated astrophysical and instrumental noise, and software influence. We assume the driving red noise patterns are correlated on one-night time scales, and are to first order uncorrelated over timescales of multiple days. Therefore, red noise of measurements on different days (or different transits) average out, and we get the total noise
\begin{align}
\sigma_\mathrm{tot}^2 = \frac{\sigma_\mathrm{white}^2 / N_\mathrm{exp} + \sigma_\mathrm{red}^2}{N_\mathrm{tr}}
\end{align}
for a phase-folded lightcurve with $N_\mathrm{tr}$ transit events.

We require two criteria for the detection of a transit signal. First, at least three transit events must be visible, $N_\mathrm{tr} \geq 3$ (section~\ref{sss:Transit Visibility}). Second, the signal-to-noise ratio, SNR, of the phase-folded lightcurve must exceed a minimum requirement. We refer to this minimum SNR as the \textit{detection threshold}, in the following denoted by the acronym $\mathrm{DT}$:
\begin{align}
SNR = \frac{\delta_\mathrm{tr/occ}} {\sigma_\mathrm{tot}} > \mathrm{DT} .
\end{align}
In here, $\delta_\mathrm{tr/occ}$ denotes the depth of the transit or occultation signal.

\subsection{Ruling out false positives}
\label{ss:Ruling out false positives}
Signals caused by eclipsing binaries are the most common astrophysical false positives in wide-field surveys for transiting planets \citep[][review]{Cameron2012}. 
To identify this configuration we define criteria based on the work of the Kepler team (\citealt{Batalha2010} and \citeyear{Batalha2012}, \citealt{Bryson2013}), which can potentially be implemented in any survey's pipeline. 
We consider that a transit signal event originates from a false positive when
\begin{itemize}
\item we measure transit depths greater than a given threshold ($2~\mathrm{R}_\mathrm{Jup}$), assuming the radius of the host star is known.
\item we detect a secondary eclipse $\delta_\mathrm{occ}$ and clearly distinguish it from the transit signal $\delta_\mathrm{tra}$ if both criteria are met:
\begin{align}
\frac{\delta_\mathrm{occ}}{\sigma_\mathrm{occ}} > \mathrm{DT} \quad \mathrm{and} \quad \frac{\delta_\mathrm{tra}  - \delta_\mathrm{occ}}{\sqrt{\sigma_\mathrm{tra}^2 + \sigma_\mathrm{occ}^2}} > \mathrm{DT}.
\end{align}
As introduced in section \ref{ss:Transit detection}, DT denotes the detection threshold, and $\sigma_\mathrm{tra}^2$ and $\sigma_\mathrm{occ}^2$ the noise of the transit and occultation signals.
\item there are ellipsoidal variations in their lightcurve, a typical feature of close binaries. We use the criteria from \cite{Sullivan2015} for the simulations of the TESS yield and employ the model of photometric variations from \cite{Mazeh2008}, using limb darkening from \cite{Claret2012} and \cite{Claret2013} as well as gravity darkening from \cite{Lucy1967} to calculate the signal caused by ellipsoidal variations. 
\item their in-/egress time equals the transit duration and the transit depth is less than $10\%$ reduced by dilution, such that the V-shape remains clearly detectable in the lightcurve. Given that planet transits can be V-shaped as well, this criteria can not be used alone (see section~\ref{ss:False positives and false negatives}).
\item if during their eclipse the center of flux in the aperture (centroid) shifts more than a given fraction of a pixel.
This aims to identify background eclipsing binaries.
\item their transit duration is significantly different than what is expected for a planet. For this purpose we calculate the theoretical transit duration of a gas giant planet ($2~\mathrm{R}_\mathrm{Jup}$) that orbits the target star with the detected period. We approximate the orbit to be circular, which is justified for short-period planets. The impact parameter $b$ dictates the transit duration and is unknown. However, it is possible to estimate a maximum transit duration by setting $b=0$. We identify whether the detected transit duration is greater than this maximum value.
\end{itemize}

\subsection{Predicted radial velocity amplitudes}
\label{ss:Doppler follow-up}
Assessing the feasibility of radial velocity follow-up for transit surveys is important to anticipate follow-up strategies.
We assume a radius versus mass relationship following \cite{Weiss2014} for objects below $3~\mathrm{R}_{\oplus}$. For planets of $\mbox{3-6}~\mathrm{R}_{\oplus}$ we adopt a Neptune density of $\rho_\mathrm{Neptune}=\SI{1.64}{g/cm^3}$. 
For $\mbox{6-11}~\mathrm{R}_{\oplus}$ we adopt a Jupiter density of $\rho_\mathrm{Jupiter}=\SI{1.33}{g/cm^3}$, and for larger planets a Jupiter mass of $\mathrm{M}_\mathrm{Jup}=\SI{1.898e27}{kg}$ as a mean value.
To reflect the intrinsic diversity of planetary composition and structure we distribute the masses following a log-normal distribution with deviation $0.5$ around the mean.
Finally, we estimate the radial velocity semi-amplitude using the parameters assigned to the planet systems and compare to the limits of current facilities.


\section{Verifying the simulation on the example of Kepler}
\label{s:Verifying the simulation on the example of Kepler}

Using our simulations, we estimate the yield of Kepler and compare our results to the Kepler candidates and confirmed Kepler planets.
We use two approaches: 1) we draw the target stars from version $10$ of the Kepler Input Catalog \citep{Brown2011}, and use the TRILEGAL galaxy model to simulate background stars and distribute them randomly in the Kepler field of view; 2) we use solely TRILEGAL and create an ad-hoc target list from all FGKM stars brighter than $V=15$. In both cases the CCD parameters and noise levels of Kepler are adopted from \cite{Gilliland2011} and the Kepler bandpass from \cite{Koch2010}.

We find that the two approaches are consistent in their yield predictions. In both cases we obtain a total of ${\sim}3000$ planets to be discovered with Kepler. These comprise ${\sim}600$ Earths, ${\sim}1000$ Super-Earths, ${\sim}1000$ Small Neptunes, ${\sim}100$ Large Neptunes, and ${\sim}200$ giant planets. 
Currently there are ${\sim}4700$ objects listed as Kepler candidates, out of which ${\sim}2300$ have so far been confirmed as planets\footnote{\url{http://exoplanetarchive.ipac.caltech.edu/} (17 Aug 2016)}.

When comparing the statistics of Kepler candidates and confirmed Kepler planets with the simulated yield we find a good agreement. 
First, the total number of simulated planets agrees with the actual findings. Second, the balance of planet types is in agreement with the statistics drawn from both the Kepler candidates as well as the confirmed Kepler planets. 
Hence, our yield results for Kepler verify our models and assumptions in the simulations. The code solely requires changing a set of priors to be used for other transit surveys. These priors contain the target list, telescope bandpass, noise levels, as well as the observation window and strategy.

\section{Estimating the yield of NGTS}
\label{s:Estimating the yield of NGTS}

\subsection{NGTS facility, target list and background stars}
NGTS is based at the European Southern Observatory's Paranal Observatory in Chile.
The facility is made of twelve fully-robotic $20$~cm telescopes with a $7.4$~sq.deg. field of view each and can be operated independently. Each CCD is a deep depleted $2\mathrm{k}\times 2\mathrm{k}$ Ikon-L produced by Andor, with pixel size of $13.5~\mathrm{\mu m}$ ($\SI{4.97}{arcsec}$). The telescopes have a constant PSF FWHM of $12~\mathrm{\mu m}$ across the field of view. More details may be found in \cite{Chazelas2012} and \cite{Wheatley2013}.

In our yield simulation we consider the situation where each telescope observes a separate neighboring field, such that the total field of view is $\SI{88.8}{sq deg}$ combined (hereafter called \textit{NGTS-field}). 
We assume the survey covers three different NGTS-fields per year within four years operation of the mission. In total $12$ NGTS-fields will be observed.
In Paranal, in average $\SI{78}{\percent}$ of the night time is of photometric quality. 
A typical night lasts $\SI{10.5}{h}$ during winter and $\SI{7.5}{h}$ during summer\footnote{\url{https://www.eso.org/sci/facilities/paranal} (17 Aug 2016)}. 
Considering the observation duration of an NGTS-field is four months and detectable transit periods are less than two weeks, we may assume all phase-folded light curves will be randomly uniformly sampled.
We select a typical NGTS-field at $l = \SI{285}{\degree}, b=+\SI{20}{\degree}$, which is representative of the targeted stellar population.
The simulation input file we employ to model NGTS can be found in appendix~\ref{app:Input file for the yield simulation on the example of NGTS}.

\begin{table}
 \caption[NGTS setting]{Settings for NGTS in the yield simulation. The target list is chosen from all FGKM stars bright enough for follow-up. The presented noise values and CCD parameters are based on observed data from both test and commissioning phases \citep[and private correspondence within the NGTS consortium]{Walker2013}.}
 \label{tab:NGTS settings}
 \center
\begin{tabular}{
l@{\extracolsep{4pt}}
l@{\extracolsep{4pt}}
l@{\extracolsep{4pt}}
l
l@{\extracolsep{4pt}}
l@{\extracolsep{4pt}}
l@{\extracolsep{4pt}}
l
}
\toprule
\multicolumn{4}{l}{Target list} & \multicolumn{4}{l}{Noise levels} \\
\midrule
$V$ & < & $15$ &  &       
$N_\mathrm{sky}^\mathrm{new~moon}$ & = & 65 & $\mathrm{e^- s^{-1} pix^{-1}}$ 	\\

$R_\star$ & < & $2$ & $\mathrm{R_\odot}$ &
$N_\mathrm{sky}^\mathrm{full~moon}$ & = & 600 & $\mathrm{e^- s^{-1} pix^{-1}}$  \\

$T_\mathrm{eff}$ & < & $10^5$ & $\mathrm{K}$ &  
$N_\mathrm{sky}^\mathrm{median}$ & = & 125 & $\mathrm{e^- s^{-1} pix^{-1}}$   \\
 
$\log g$ & <  & $6.5$	  &  &                           
$N_\mathrm{dark}$ & = & 0.06 & $\mathrm{e^- s^{-1} pix^{-1}}$  \\
 
 & & & &
$N_\mathrm{read}$ & = & 10 & $\mathrm{e^- pix^{-1}}$ \\
 
 & & & &
 $\sigma_\mathrm{red}$ & = & $\SIrange{0}{2}{}$ & $\mathrm{mmag}$ \\
 
 \midrule
 \multicolumn{4}{l}{CCD parameters} & \multicolumn{4}{l}{} \\
 \midrule
 $t_\mathrm{exp}$ & = & $10$ & $\mathrm{s}$  &
 & & & \\
 
 $t_\mathrm{read}$ & = & $1.49$ &  $\mathrm{s}$ &
 & & & \\

\bottomrule 
\end{tabular} 
\end{table}

\subsection{Red noise as the dominant limitation}
\label{ss:Red noise as the dominant limitation}
We compute the white noise using Eq.~\ref{eq:white_noise} and the parameters shown in Table~\ref{tab:NGTS settings} (see also appendix~\ref{app:Input file for the yield simulation on the example of NGTS}).
Based on the design of NGTS, we consider circular apertures with a $\SI{3}{pixel}$ radius.
The sky noise varies strongly with lunar phase; we adopt the median value during a lunar cycle.
We compute the scintillation noise using \cite{Dravins1998} for $\SI{20}{cm}$ telescopes, an average airmass of $1.5$ and location at $\SI{2400}{m}$ above sea level. 

\begin{figure}
 \includegraphics[width=\columnwidth]{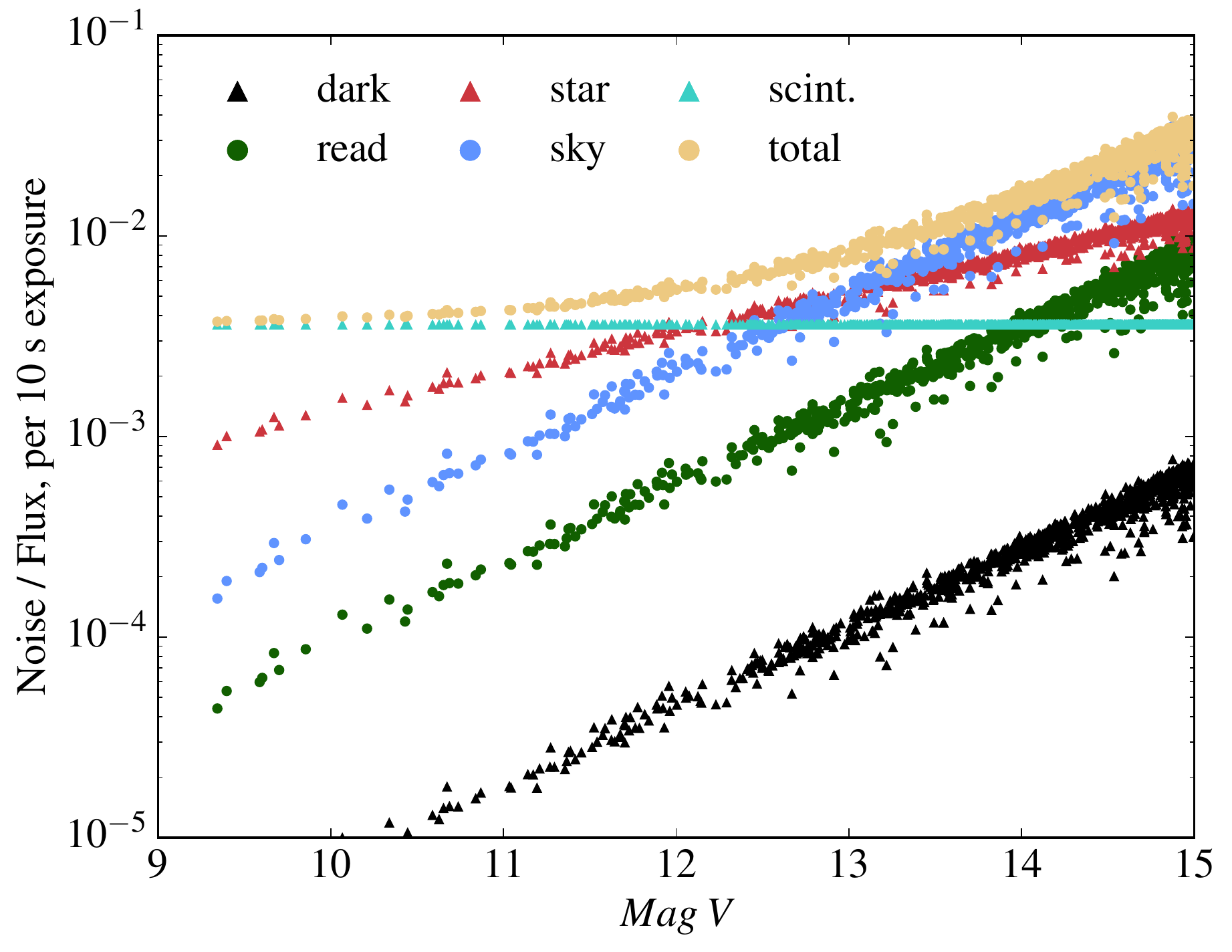}
 \caption[]{NGTS' white noise components. Values are calculated for each star in the simulated run time using Eq.~\ref{eq:white_noise}, the stellar flux (section~\ref{ss:Stars and photometry}) and empirical values for sky, dark and readout counts (Table~\ref{tab:NGTS settings}). Color versions of all figures are available in the online journal.}
 \label{fig:white noise}
\end{figure}

Scintillation noise is the main white noise component for a single exposure for stars brighter than $V=11$ (Fig.~\ref{fig:white noise}). For fainter targets, stellar and background noise become driving factors, with the latter dominating at the faint end for $V>13.5$.

\begin{figure}
 \includegraphics[width=\columnwidth]{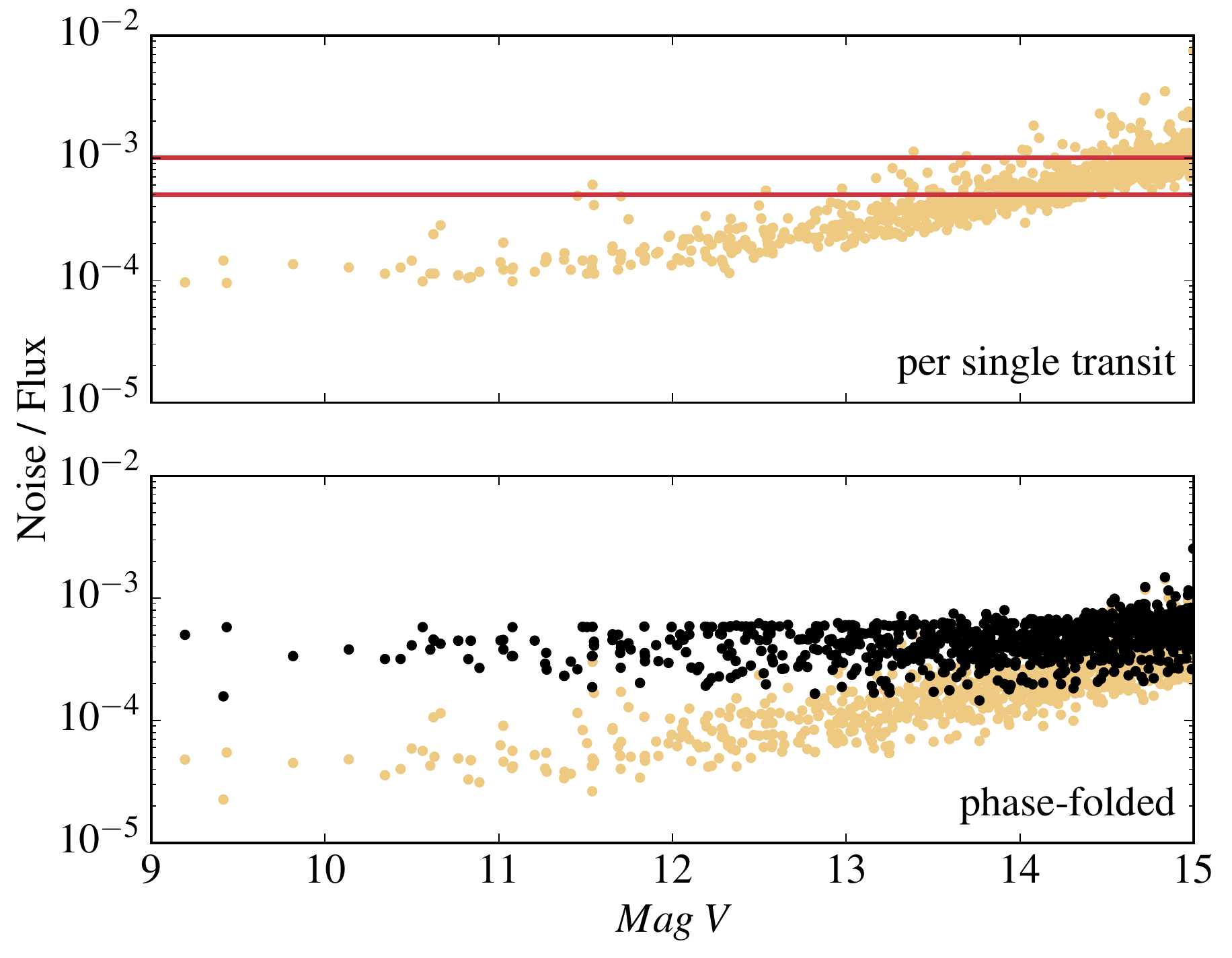}
 \caption[]{NGTS' white and red noise. Upper panel: white noise (yellow circles) binned over all exposures in one single transit with typical transit timescales of $\SIrange{1}{4}{h}$. Red lines indicate the impact of red noise levels at $\SI{0.5}{mmag}$ and $\SI{1}{mmag}$. Lower panel: white noise (yellow circles) and total noise (i.e. white noise and $\SI{1}{mmag}$ red noise; black circles) for phase-folded lightcurves.}
 \label{fig:total noise}
\end{figure}

Correlated noise and systematics, referred to as red noise, affect photometric measurements on timescales comparable to the transit duration (few hours). NGTS tests in Geneva and La Palma demonstrated its ability to achieve $\SI{1}{mmag}$ sensitivity and better \citep{Wheatley2013}.
The upper panel of Fig.~\ref{fig:total noise} illustrates the white noise binned up per single transit, $\sigma_{\mathrm{white}}/N_{\mathrm{exp}}$ (section~\ref{ss:Transit detection}). 
Considering a single-transit, a red noise level of $\SI{1}{mmag}$ dominates the total noise for objects $V<14$, a red noise of $\SI{0.5}{mmag}$ dominates for objects $V<13$. Note that for phase-folded lightcurves of bright targets a white noise model would underestimate the total noise by up to an order of magnitude compared to a model taking $\SI{1}{mmag}$ of red noise into account (Fig.~\ref{fig:total noise}, lower panel).

\subsection{Choosing a minimum detection threshold}
\label{ss:Choosing a minimum detection threshold}
The choice of the \textit{detection threshold} $\mathrm{DT}$ (introduced in section \ref{ss:Transit detection}), the minimum signal-to-noise ratio required to trigger a detection, has a significant impact on the expected yield.
A high detection threshold leads to non-detection of small planets.
On the contrary, lowering the detection threshold increases the number of false alarms. These are commonly caused by systematic errors referred to as red noise. The impact of red noise depends on the timescale we consider. 
In this section we estimate the optimal detection threshold for NGTS as a function of the estimated number of false alarms.

To estimate the number of false alarms, we begin our argument with a time series of $N_\mathrm{tot}$ data points in which each point corresponds to an average of the measurements taken over a duration equal to the a typical transit duration ($\SIrange{1}{4}{h}$). The chance that Gaussian noise causes one data point to lie off the mean (looking like a transit) by a standard deviation more than $x~\sigma$ is given by
\begin{align}
\label{eq:false_alarm_p}
p_\mathrm{outlier}\left( x \right ) &= \frac{{1 - \mathrm{erf}(x/\sqrt{2})}}{2},
\end{align}
whereby $\mathrm{erf}$ denotes the error function.
We assume there is a number of $N>3$ outliers in the time series. In order to mimic a transit pattern these $N$ outliers must be distributed in a time periodic manner. The probability for such a configuration is given by
\begin{align}
\label{eq:false_alarm_pN}
p_\mathrm{N~outliers} =
\left( p_\mathrm{outlier}(x) \cdot \frac{N_\mathrm{tot}}{N} \right)^2
\cdot
p_\mathrm{outlier}(x)^{N-2}
\quad (N>3).
\end{align}
In here, the first term ($p_\mathrm{outlier}(x) \cdot \frac{N_\mathrm{tot}}{N}$) corresponds to the probability of the first and last outlier. 
The time periodicity is defined by the total number of outliers and the positions of the first and last outlier in the time series. The outliers inbetween consequently have determined locations in the time series and the probability of this to happen is expressed by the second term of the equation.
For example, if $N=3$ the first outlier may be located anywhere in the first third of the time series ($N_\mathrm{tot}/3$ possible locations), while the last outlier may be located anywhere in the last third of the time series. The second outlier has to lie exactly in the middle between the first and last outlier.

For NGTS, the total error on each data point is dominated by red noise (see section~\ref{ss:Red noise as the dominant limitation}). Assuming red noise is uncorrelated on multi-day timescales and follows a Gaussian error distribution, it averages out with the number of transit events, and we can approximate $x \approx \mathrm{DT} / \sqrt{N}$. 
For example, with a detection threshold $\mathrm{DT}=5$ and requiring $N=3$ outliers, from Eq.~\ref{eq:false_alarm_p} follows that $p_\mathrm{outlier}\left( x \right ) \approx \SI{1.9e-3}{}$. 
In our estimation for NGTS we have $t_\mathrm{total} \approx \SI{885}{h}$. 
Assuming transit-like timescales $T \approx \SI{2}{h}$ this leads to $N_\mathrm{tot}\approx 440$.
Finally, with the example of $N=3$ outliers we obtain from Eq.~\ref{eq:false_alarm_pN} that $p_\mathrm{N = 3~outliers} \approx \SI{1.7e-4}{}$.

The false alarm probability for one time series of data points is then the sum over $p_\mathrm{N~outliers}$ for all possible $N$, which happens to converge quickly with increasing $N$.
As a false alarm can be triggered for each object in the target list, the total number of false alarms, $N_\mathrm{FA}$, scales with the number of objects in the observed field, $N_\mathrm{obj}$, and the number of covered NGTS-fields, leading to
\begin{align}
\label{eq:false_alarms}
N_{FA} &= \sum_\mathrm{fields} \left[ \left( \sum_{N=3}^{N_{max}} p_\mathrm{N~outliers} \right) \cdot N_\mathrm{obj} \right].
\end{align}

We compute Eq.~\ref{eq:false_alarms} for a range of typical transit-like timescales from $T=\SIrange{1}{4}{h}$ to evaluate the impact of the detection threshold on the yield versus false alarms.
With a total of $12$ NGTS-fields, each containing $N_\mathrm{obj} \sim 10^5$, $\mathrm{DT}=5$ leads to a number of false alarms on the order of $10^2$ among the planet candidates. 
This suggests a detection threshold of at least $\mathrm{DT}=5$ should be used.

\subsection{Expected yield and major influencing factors}
\label{ss:Expected yield and major influencing factors}

\subsubsection{A multitude of Neptunes and giants}
\begin{figure}
 \includegraphics[width=\columnwidth]{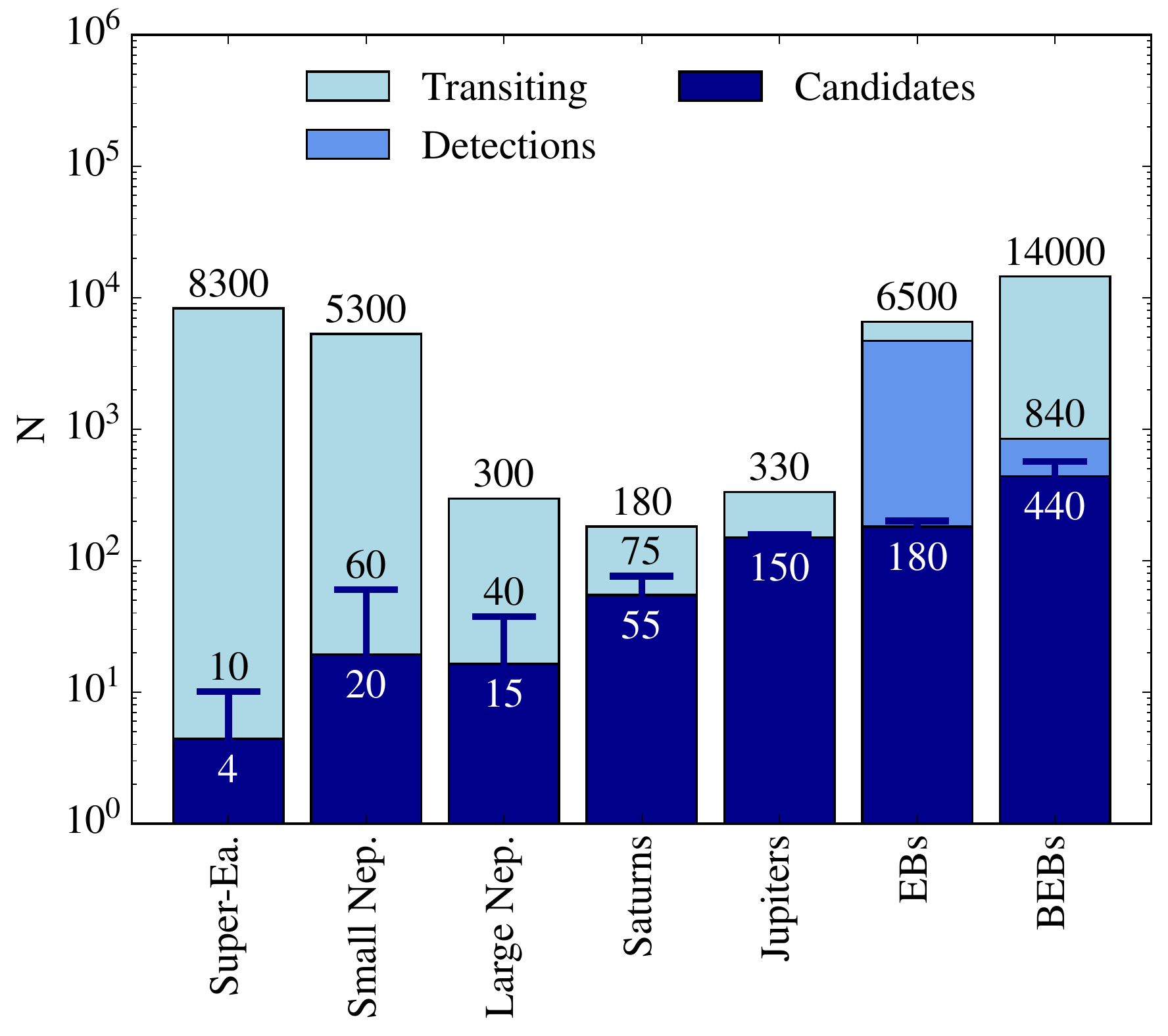}
 \caption[]{Expected Yield for NGTS' planets (see Table~\ref{tab:planet_classes}) and false positives, i.e. eclipsing binaries (EBs) and background eclipsing binaries (BEBs). Light blue: objects undergoing a transit in the line of sight with orbital periods shorter than $\SI{20}{days}$. Blue: Objects that can be detected with a detection threshold $\mathrm{DT}=5$ for a red noise level of $\SI{1}{mmag}$. Dark blue: planetary candidates that remain from the former group after applying the rule-out criteria for false positives described in section~\ref{ss:Ruling out false positives}. These remaining false positives in the NGTS planet candidate list need follow-up with additional instruments before they can be identified. Blue lines indicate the possible yield if a red noise level of $\SI{0.5}{mmag}$ can be reached. All values are averaged over ten simulation runs. Uncertainties are indicated in Table~\ref{tab:expected yield}.}
 \label{fig:yield_histogram}
\end{figure}

We categorize the yield by object categories: eclipsing binaries (EB), background eclipsing binaries (BEB) and planet types (shown in Table~\ref{tab:planet_classes}). For the application to NGTS we subdivide the category of giant stars from section \ref{sss:Planets} into Saturns ($\mbox{6-10}~\mathrm{R}_{\oplus}$) and Jupiters ($\mbox{10-22}~\mathrm{R}_{\oplus}$).
The simulation over the entire survey time of four years is repeated ten times using the same parameters and input to estimate the mean values and standard deviations of the total number of expected planets and false positives (Fig.~\ref{fig:yield_histogram}).
We account for uncertainties in the planet and binary priors via error propagation. 
Statistical errors caused by re-running the TRILEGAL galaxy model for the same field are found to be negligible in comparison.

\begin{table}
\centering
 \caption[Planet Classes]{Planet classification based on the radius.}
 \label{tab:planet_classes}
 \begin{tabular}{
 			 	l
 				S[table-format=2.2]@{\extracolsep{4pt}}
 				l@{\extracolsep{4pt}}
			    S[table-format=2]@{\extracolsep{4pt}}
			    l@{\extracolsep{4pt}}
 			}
  \toprule
  Super-Earths & 1.25 & -- & 2 & $~\mathrm{R_{\oplus}}$\\
  Small Neptunes & 2 & -- & 4 & $~\mathrm{R_{\oplus}}$\\
  Large Neptunes & 4  & -- & 6 & $~\mathrm{R_{\oplus}}$\\
  Saturns & 6 & -- & 10 & $~\mathrm{R_{\oplus}}$\\
  Jupiters & 10 & -- & 22 & $~\mathrm{R_{\oplus}}$\\
  \bottomrule
 \end{tabular}
\end{table}

NGTS' combined fields over four years contain tens of thousands of planets and binary systems that transit in the line of sight (Fig.~\ref{fig:yield_histogram}). After accounting for visibility, noise, and detection criteria, only a small fraction triggers a detectable signal. A detection threshold $\mathrm{DT}=5$ for a $\SI{1}{mmag}$ red noise leads to the detection of ${\sim}35$ small and large Neptunes as well as ${\sim}200$ of giant planets.

\subsubsection{Impact of red noise and detection criteria}

The impact of the red noise level and detection threshold is significant for Neptune-sized planets (Table~\ref{tab:expected yield}). 
If we decrease the red noise from $\SI{1}{mmag}$  to $\SI{0.5}{mmag}$ , the sensitivity for small planets increases by a factor of three (see also Fig.~\ref{fig:yield_histogram}). 
Omitting red noise leads to $\sim\SIrange{250}{300}{}$ additional small planets.
Increased red noise levels lead to a loss of most small planets.

Maintaining the assumption of a $\SI{1}{mmag}$  red noise and reducing the detection threshold from $\mathrm{DT}=5$ to $3$ leads to a comparable increase as dividing the red noise by two, but increases the number of false alarms by several magnitudes as discussed in section~\ref{ss:Choosing a minimum detection threshold}. In contrast, increasing the detection threshold from $\mathrm{DT}=5$ to $7$ leads to the loss of small planets comparable to doubling the red noise.

For the purpose of the following sections we assume a detection threshold $\mathrm{DT}=5$ and, confirm with the design goal of NGTS, a $\SI{1}{mmag}$ red noise unless otherwise stated.

\begin{table}
\centering
 \caption[Expected Yield]{Expected yield for NGTS for different assumptions of the red noise. The table shows all planet classes, false positives before the vetting process described in section~\ref{ss:Ruling out false positives} (EB before, BEB before), and false positives that remain unidentified after the vetting process (EB after, BEB after). Shown values are the mean and standard deviation averaged over $10$ simulation runs, including any uncertainties on the priors.}
 \label{tab:expected yield}
\begin{tabular}{
				 l
				 S[table-format=4]@{\,\( \pm \)\,}
			     S[table-format=2]
   				 S[table-format=4]@{\,\( \pm \)\,}
   				 S[table-format=2]
   				 S[table-format=4]@{\,\( \pm \)\,}
   				 S[table-format=2]
   				 S[table-format=4]@{\,\( \pm \)\,}
   				 S[table-format=2]
                }
 \toprule
 & \multicolumn{2}{r}{$\SI{2}{mmag}$}  & \multicolumn{2}{r}{$\SI{1}{mmag}$}  & \multicolumn{2}{r}{$\SI{0.5}{mmag}$}  & \multicolumn{2}{r}{$\SI{0}{mmag}$}  \\ 
 \midrule
Super-Earths & 1 & 1 & 4 & 3 & 10 & 3 & 28 & 3 \\ 
Small Nep. & 5 & 2 & 19 & 5 & 60 & 10 & 229 & 20 \\ 
Large Nep. & 4 & 1 & 16 & 4 & 38 & 4 & 69 & 6 \\ 
Saturns & 28 & 5 & 55 & 8 & 76 & 10 & 86 & 10 \\ 
Jupiters & 129 & 9 & 150 & 10 & 158 & 11 & 161 & 10 \\ 
 \midrule
EB before & 4719 & 45 & 4688 & 45 & 4708 & 46 & 4719 & 45 \\ 
BEB before & 1070 & 88 & 843 & 75 & 972 & 83 & 1070 & 88 \\ 
EB after & 211 & 12 & 181 & 12 & 201 & 12 & 211 & 12 \\ 
BEB after & 665 & 51 & 439 & 37 & 568 & 47 & 665 & 51 \\ 
  \bottomrule
 \end{tabular}
\end{table}

\subsection{False positives and false negatives}
\label{ss:False positives and false negatives}

NGTS finds a significant number of false positives, consisting of ${\sim}4700$ EBs and ${\sim}850$ of BEBs (Table~\ref{tab:expected yield}, Fig.~\ref{fig:yield_histogram}). 
To copy the screening process to identify them, we use the series of criteria described in section~\ref{ss:Ruling out false positives}. 
We only vet target stars for ellipsoidal variations, assuming lightcurve features of diluted background stars are undetectable with NGTS. The centroiding sensitivity for NGTS is set to $1/100$ pixel.

\begin{figure}
 \includegraphics[width=\columnwidth]{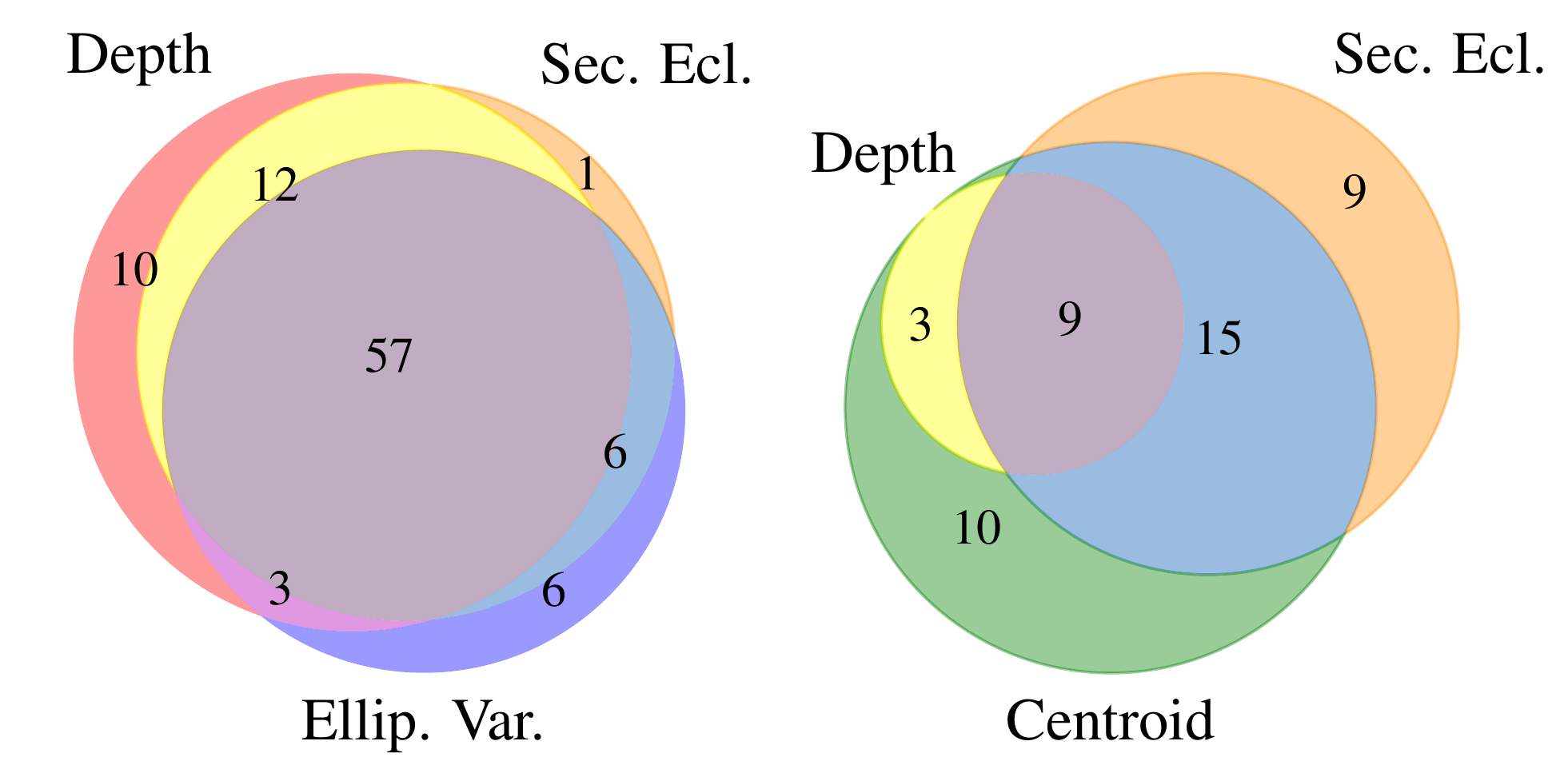}
 \caption[]{Percentage of false positives identified with different rule-out methods for EBs (left) and BEBs (right). 
 Overlapping regions are designated by color combination. The relative error on all values is $\leq \SI{10}{\percent}$.}
 \label{fig:FP_ruleout_venn}
\end{figure}

The most efficient criteria are depth, secondary eclipses and ellipsoidal variations for EBs, as well as centroiding and secondary eclipses for BEBs (Fig.~\ref{fig:FP_ruleout_venn}). Overall we estimate $\SI{96}{\percent}$ of EBs and $\SI{48}{\percent}$ of BEBs can be identified by the vetting process.
These values include vetting for the transit duration, which can be used to identify $\SI{5}{\percent}$ of false positives, including $\SI{2}{\percent}$ that can not be detected with another method. V-shaped transits can be detected for $\SI{54}{\percent}$ of EBs, including $\SI{5}{\percent}$ that can not be ruled out with other methods. Here, we do not use the V-shape alone to reject an object, as planets can cause V-shaped transits as well.

While individual criteria can already give a hint towards possible false positives, at least two criteria can be met in parallel for about three quarters of all EBs and one quarter of all BEBs.
All criteria are effective for a wide range of EBs, but are less applicable for systems with faint secondaries. 
Faint BEBs are strongly diluted, which decreases the ability to identify them with any method.

The search for false positives can lead to false negatives.
Giant planets undergoing grazing eclipses can still trigger detectable NGTS signals. We estimate $\SI{10}{\percent}$ of all planet detections show a distinguishable V-shape.
The transit depth is a reliable measure assuming the stellar properties of the target star are known well.
Secondary eclipses, ellipsoidal variations and centroid shifts for planets are not expected to be detectable with NGTS.

\subsection{Characteristics of NGTS candidates}

\subsubsection{Distinguishing remaining false positives from planets}

\begin{figure}
 \includegraphics[width=\columnwidth]{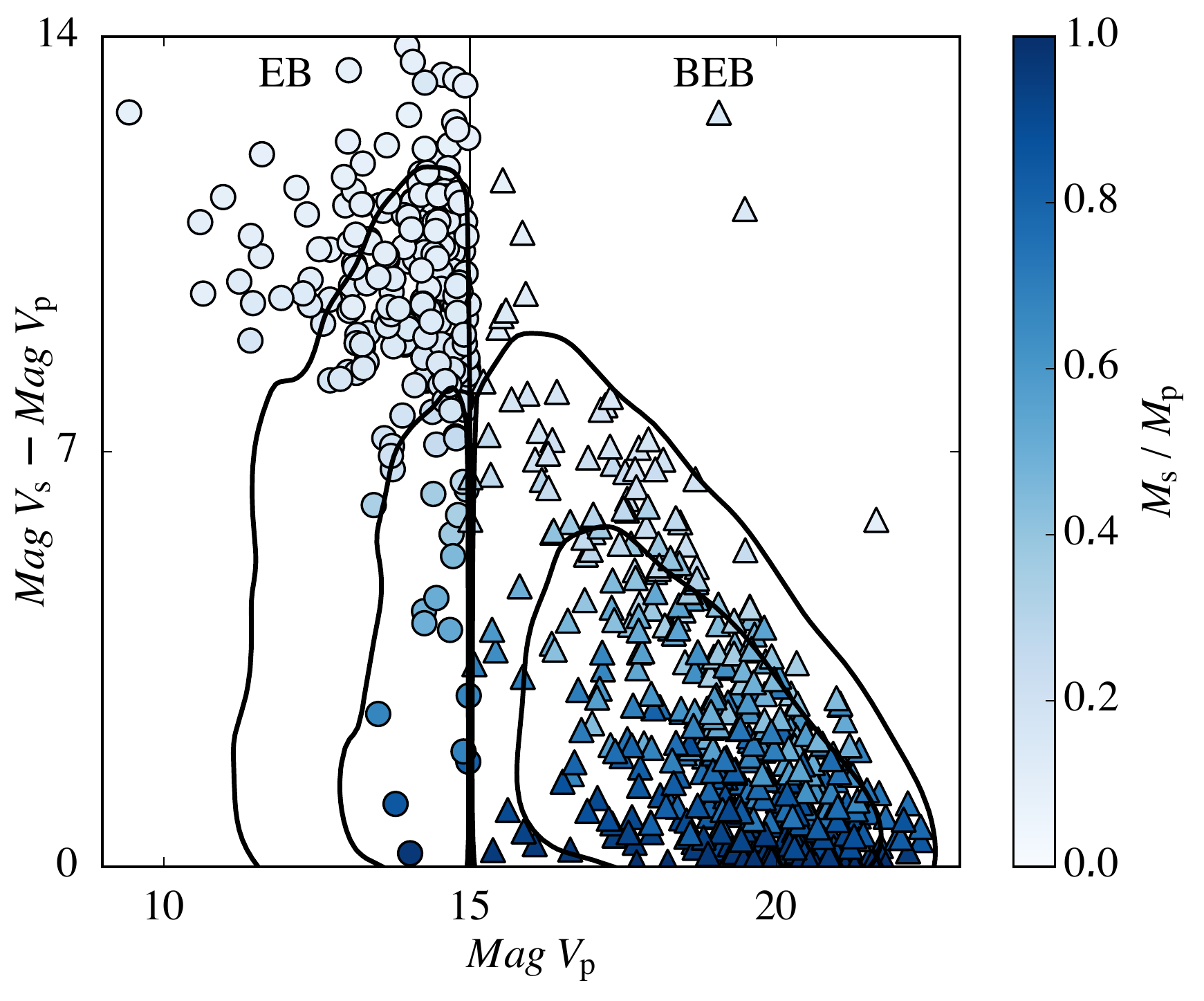}
 \caption[]{EBs (circles) and BEBs (triangles) that are not identified by the lightcurve vetting process and need follow-up observations, shown against the magnitudes of the primary and secondary. The color-coding displays the binary mass ratio $M_\mathrm{s}/M_\mathrm{p}$ of the secondary to primary. Contour lines illustrate the original distribution of false positives before the vetting process.}
 \label{fig:FP_remaining}
\end{figure}

\begin{figure}
 \includegraphics[width=\columnwidth]{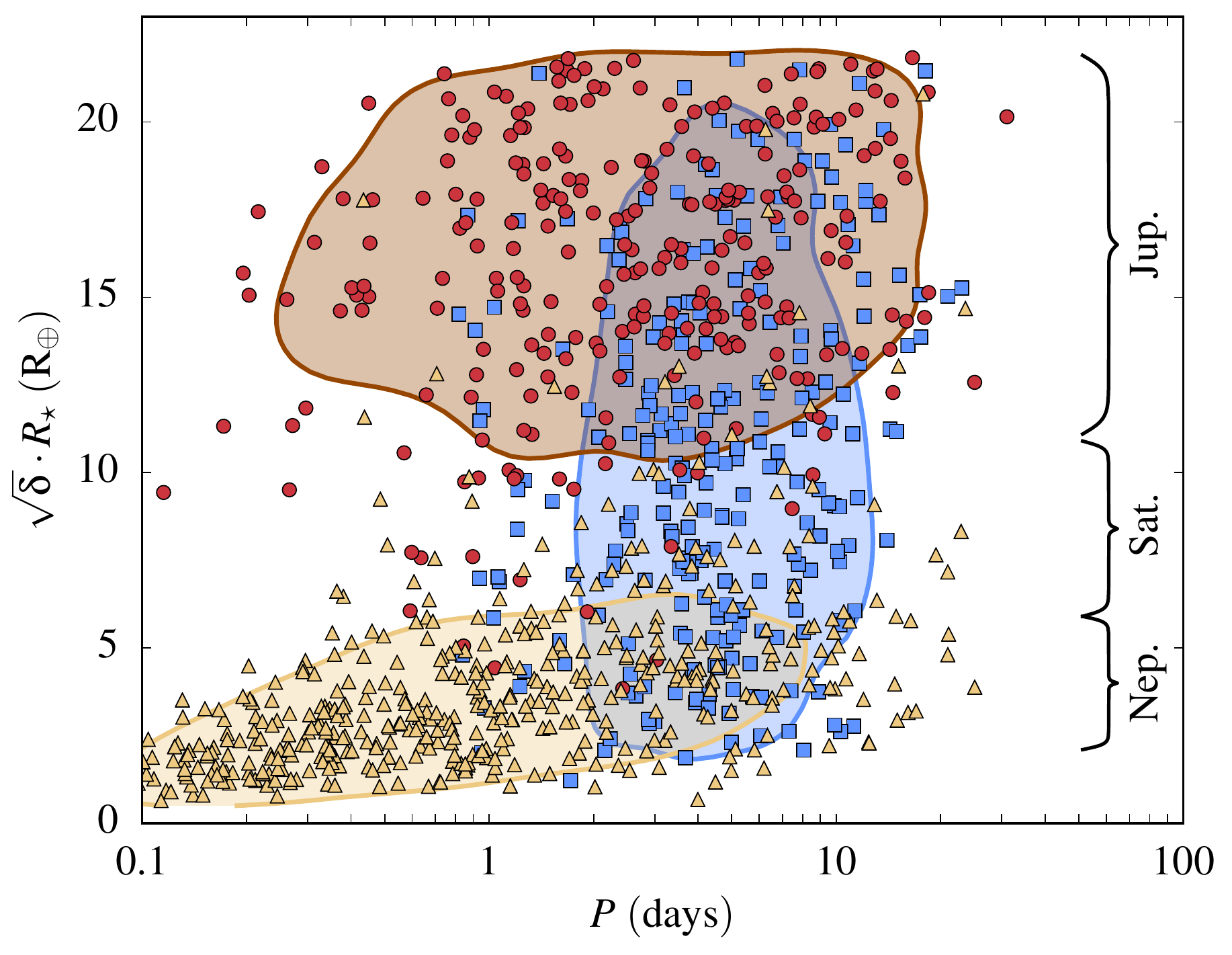}
 \caption[]{Objects in the expected NGTS candidate list after the photometric vetting process. Shown is the measured apparent radius without correction for dilution or grazing eclipses. Planets (blue squares), EBs (red circles) and BEBs (orange triangles) occupy different regions of the parameters space in apparent transit radius and orbital period. Symbols denote the objects in one simulation run for a red noise of $\SI{1}{mmag}$, $1~\sigma$ contours are calculated over ten simulation runs.}
 \label{fig:planet_fp_Rp_P}
\end{figure}

After the candidate vetting process described above, the majority of false positives will be identified but the undetected ones may still outnumber the planet candidates. EBs that remain undetected in the candidate list are expected to consist of binaries with low mass companions, such as M-stars or Brown Dwarf secondaries (Fig.~\ref{fig:FP_remaining}). 
Small M stars or Brown Dwarfs are the same size as gas giants \citep{Fortney2011} and hence pollute the sample of giant planets but scarcely affect the population of smaller planets (Fig.~\ref{fig:planet_fp_Rp_P}).
These systems cannot be ruled out based on a transit lightcurve, but need radial velocity follow-up and mass measurements.

Remaining BEBs can show more variety due to different degrees of dilution of their transit signals (Fig.~\ref{fig:FP_remaining}). They can mimic planetary radii over a large range but especially pollute the sample of planets in the Neptune-sized regime (Fig.~\ref{fig:planet_fp_Rp_P}).
In addition, this is the regime where statistical false alarms will pollute the sample most.

Fig.~\ref{fig:planet_fp_Rp_P} further illustrates that it will be difficult to filter out Planets below $2$-day orbital periods.
The planet occurrence rates suggest planets are unlikely to have orbital periods of less than $\SIrange{1}{2}{days}$ \citep{Fressin2013, Dressing2015}. Many binary systems, in contrast, are known to orbit on time scales of only a few hours, including detached systems \citep[see e.g.][]{Norton2011,Soszynski2015}.

\subsubsection{Expected planet properties}

\begin{figure}
 \includegraphics[width=\columnwidth]{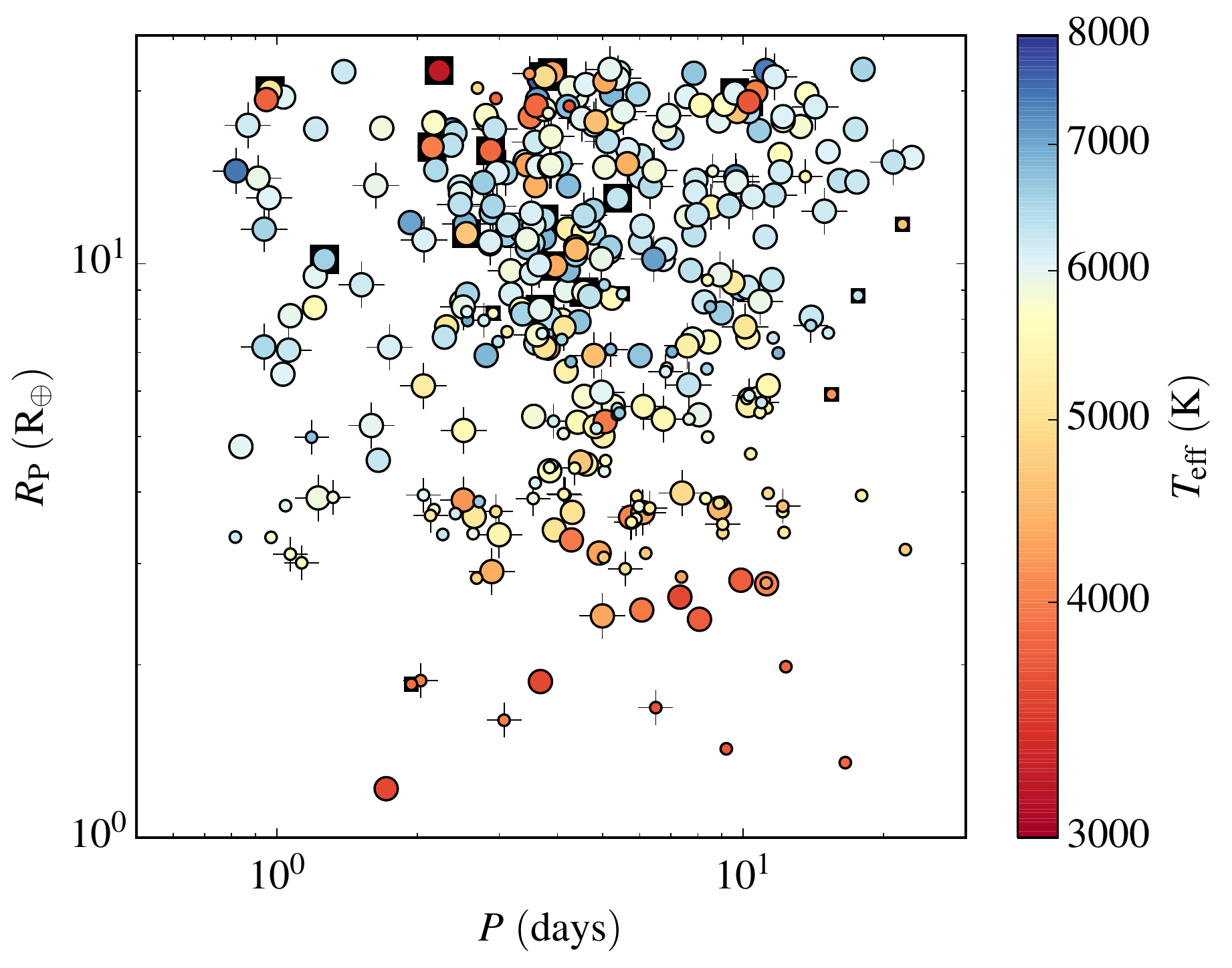}
 \caption[]{Distribution of detected NGTS planets in period and radius, color-coded in dependence of the host star's effective temperature. Symbol sizes represent detections for red noise levels of $\SI{0.5}{mmag}$ (small) and $\SI{1}{mmag}$ (large). An underlying cross (square) marks a planet orbiting the primary (secondary) of a binary system.}
 \label{fig:planets_hosts_P}
\end{figure}

\begin{figure}
 \includegraphics[width=\columnwidth]{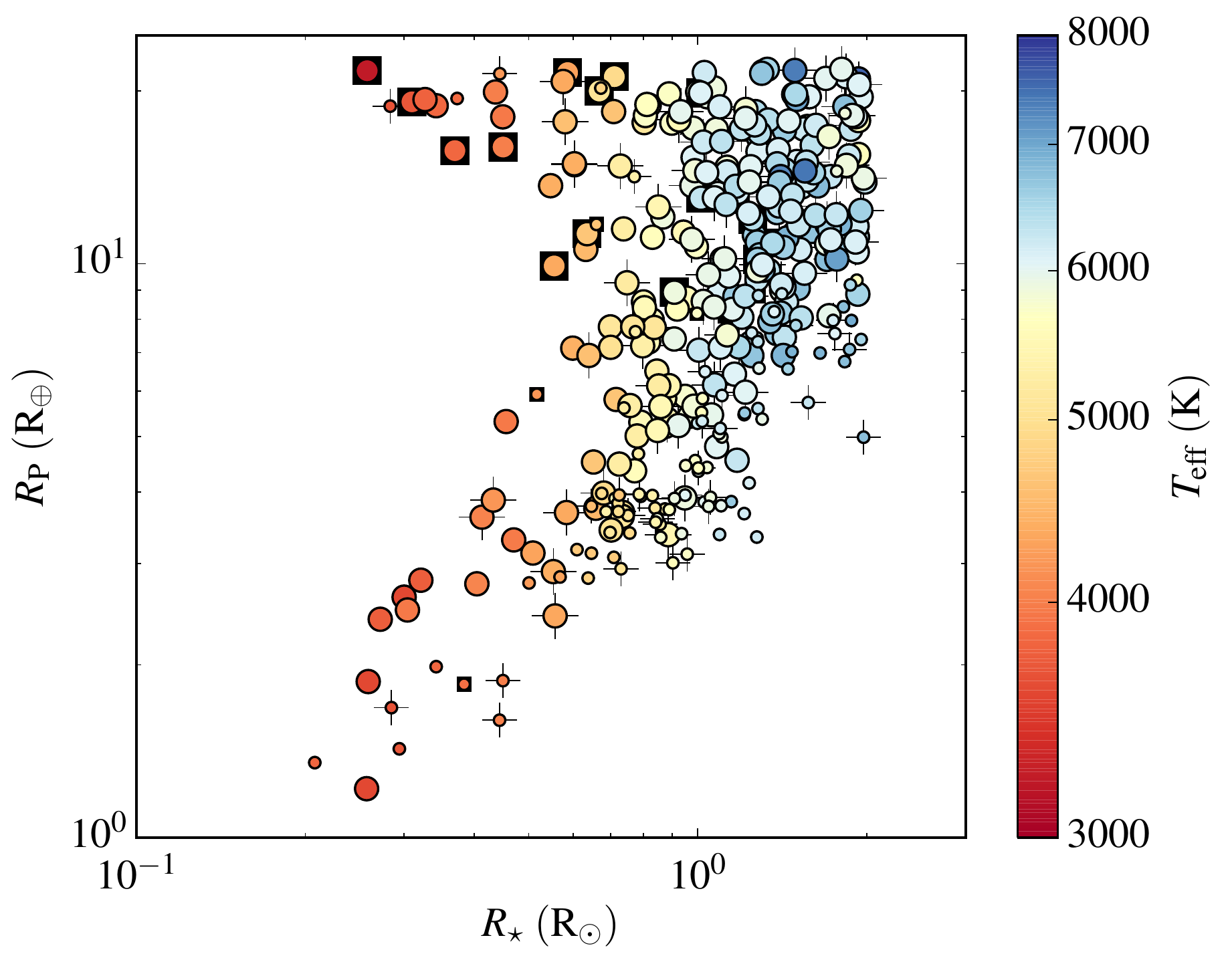}
 \caption[]{Radius of detected NGTS planets versus radius of their host star, color-coded by the host star's effective temperature. Symbol sizes represent detections for red noise levels of $\SI{0.5}{mmag}$ (small) and $\SI{1}{mmag}$ (large). An underlying cross (square) marks a planet orbiting the primary (secondary) of a binary system.}
 \label{fig:planets_hosts_Rs}
\end{figure}

$97\pm1\%$ of all predicted NGTS planets are found at orbital periods shorter than two weeks, and $60\pm3\%$ orbit at less than $\SI{5}{days}$ (Fig.~\ref{fig:planets_hosts_P}).
Decreasing the red noise by half leads to an increase of NGTS' sensitivity to detect small planets with longer orbital periods.
K stars are found to be typical hosts for NGTS' smallest planets, such as Small Neptunes and Super-Earths (Figs.~\ref{fig:planets_hosts_P} and \ref{fig:planets_hosts_Rs}). 
Large Neptunes can be detected around G stars.
Fig.~\ref{fig:planets_hosts_Rs} further illustrates the sensitivity cutoff of NGTS as a function of planet over stellar radius. The paucity of giant planets detected around small stars is a direct consequence of the planet occurrence rates estimated from the results of the Kepler mission \citep{Fressin2013,Dressing2015}.

We estimate that $57\pm17\%$ of planets detected with NGTS orbit single stars, and $34\pm12\%$ ($8\pm5\%$) orbit the primary (secondary) stars of binary systems. 
Most of the planets detected in binary systems are Jupiters and inflated giants, as only large planets are still detectable given the strong dilution of the transit signal by the light of the binary companion. 
Dilution decreases the transit depth by $\SI{20}{\percent}$ or more for $\SI{21}{\percent}$ of Jupiters, $\SI{11}{\percent}$ of Saturns and $\SI{8}{\percent}$ of smaller planets.
Circumbinary planets are not detected due to the limited sensitivity of NGTS for long-period transiting planets.

$20\pm5$ planets are detected with NGTS around background stars. These consist of systems with a faint target star and a small background star orbited by an inflated giant planet. 

\subsubsection{Radial velocity follow-up and characterization of NGTS planets}

\begin{figure}
 \includegraphics[width=\columnwidth]{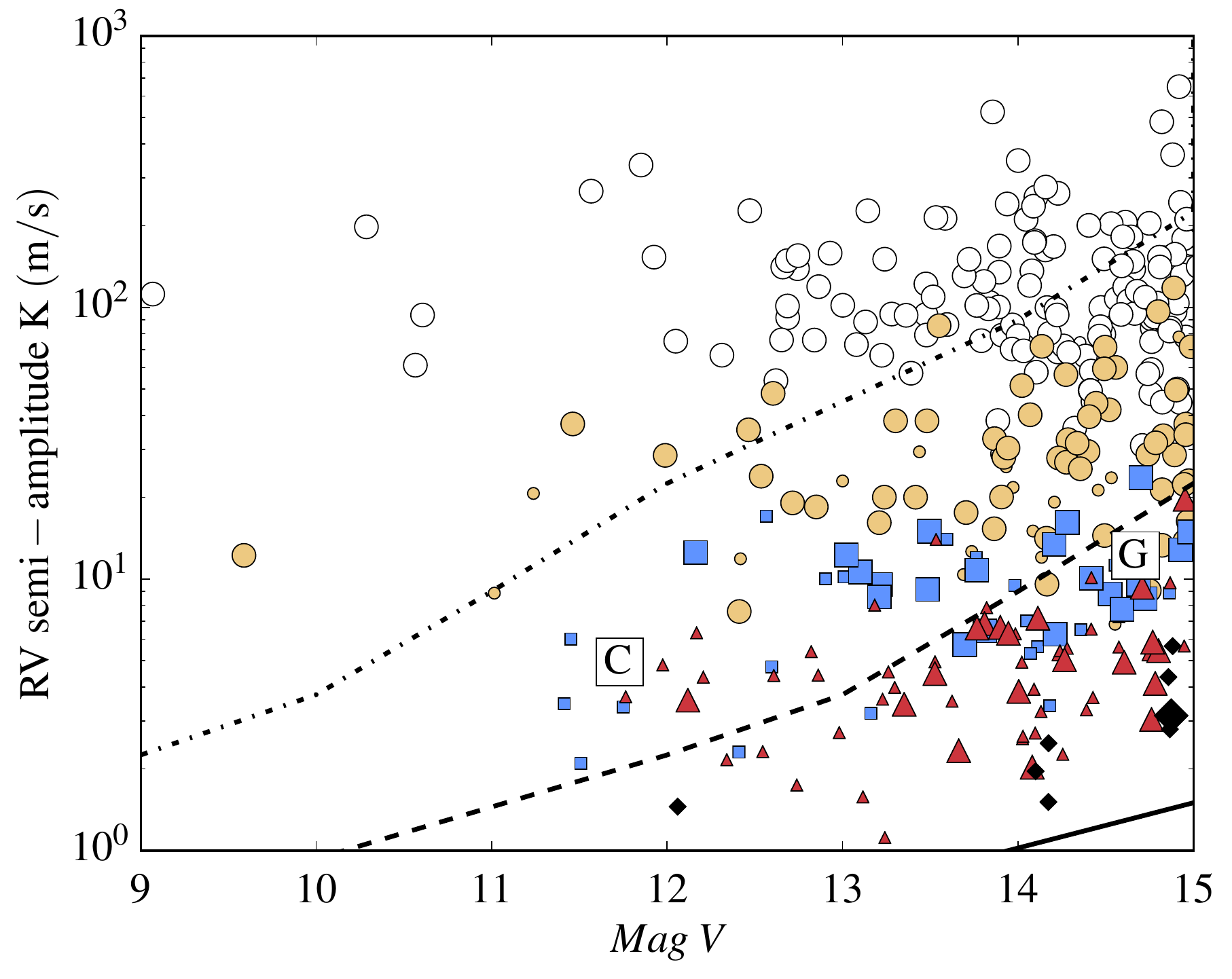}
 \caption[]{Estimated RV signals of NGTS planets compared with sensitivity of current and future RV instruments: Coralie (dotted line), HARPS (dashed line) and ESPRESSO (solid line). Planet symbols denote Jupiters (white circle), Saturns (yellow circle), Large Neptunes (blue squares), Small Neptunes (red triangles) and Super-Earths (black diamonds). Symbol sizes represent detections for red noise levels of $\SI{0.5}{mmag}$ (small) and $\SI{1}{mmag}$ (large). The sensitivity curve for each instrument corresponds to the criterion $2 K > 3~\sigma_{RV}$. This reflects the minimum signal that can be detected with $3~\sigma$ for an $\SI{1}{h}$ exposure considering the instrument's sensitivity, implying at least 10 independent RV measurements per target. The ESPRESSO sensitivity is taken from \citep{Pepe2014}, values for HARPS and Coralie are established for $\SI{1}{h}$ exposures on non-rotating and non-active stars compiled from published results. The planets Corot7b (C) and GJ 1412b (G), for which masses were measured with HARPS, are shown for comparison.}
 \label{fig:planets_MagV_k}
\end{figure}

We estimate the planetary masses and radial velocity (RV) signals following section~\ref{ss:Doppler follow-up}. 
The bulk of detected planets lies in magnitude between Corot~7b and GJ~1412b (Fig.~\ref{fig:planets_MagV_k}). Instruments like Coralie \citep{Queloz2000} are important for vetting false positives, and enable mass measurement of $43\pm2\%$ of predicted Jupiter-sized planets. 
HARPS \citep{Mayor2003} can confirm $99\pm0\%$ of all predicted Jupiter-sized, $92\pm1\%$ of Saturn-sized planets, and $51\pm2\%$ of Large Neptunes.
ESPRESSO \citep{Pepe2014} will reach the sensitivity to measure RV signals of all NGTS planets.
It is worth mentioning that most stars hosting small planets are K dwarfs (Fig.~\ref{fig:planets_hosts_Rs}) and are brighter in the infrared. Future characterization of these objects may hence be easier in that wavelength.

\section{Discussion}
\label{s:Discussion}

\subsection{Red noise limitation, detection criteria and planet merit}
Red noise is a limiting factor in searching for small planets around bright stars. According to our simulation at least a $\SI{1}{mmag}$ precision is essential to detect Neptune-sized planets (see Table~\ref{tab:expected yield}). 
In this work we assume major components of the red noise are only short-time (nightly) correlated and average out over a whole season. This assumption is based on early NGTS test data (private communication within the NGTS consortium), and extrapolated from WASP results \citep{Pollacco2006} and recent studies with EulerCam at La Silla \citep{Lendl2013}. Early results from NGTS data suggest this assumption is realistic (Weathley et al. 2016 in prep.).

Red noise is the dominant factor for bright objects (see Fig.~\ref{fig:total noise}). Decreasing it by half increases the overall number of small planets and allows to find more of them around bright stars; however, it does not significantly change the number of giant planets, as the detection of giants is limited by the observation window function rather than the noise threshold.

The detection efficiency and false alarm rate depend on the chosen detection criteria, as well as on the final version of detrending and lightcurve fitting algorithms. Assuming false alarms can well be ruled out by visual inspection, an 100:1 false alarm ratio is a realistic compromise between a high planet yield and practicality. In the simulation we settled for a detection threshold of $5~\sigma$. In systems where a transit has already been found (e.g. by the TESS survey) or other planets are known, additional transit signals with lower signal-to-noise ratio may be considered.

\subsection{Selecting the target list}
The number of transiting planets scales with the number of stars in the field of view and hence decreases with distance form the galactic plane.
Conversely, there will be more background stars, leading to increased dilution and therefore a decrease of the transit depth as well as an increased number of false positives. 
Specifically, the yield of EBs, BEBs, and giant planets scales with the crowdedness of the field, as their signals are still detectable even in a crowded environment.
In contrast, Neptunes and smaller planets show a saturation behavior since dilution limits the detection of small planets in crowded fields. Sparse fields lead to a higher detection efficiency for small planets but a decreased number of targets. The results suggest these two factors average out for Neptunes and smaller planets.
It would therefore be beneficial to stay distant from the galactic plane in order to optimize for a higher yield of Neptunes and Super-Earths while reducing the contamination by false positives.

\subsection{Follow-up and characterization of NGTS candidates}
We estimate for NGTS that the majority of EBs and BEBs can be identified without necessitating further follow-up measurements. 
EBs that can not be identified by the vetting process usually have a low-mass secondary such as an M-dwarf or Brown dwarf. They can mimic planets with sizes of Jupiter or greater. It will require RV follow-up to determine the mass of the companion and reject a planet hypothesis. 
While some of the remaining BEBs can have low-mass companions and hence mimic various planetary signals, the majority consists of binary companions with comparable mass. These systems cause very deep transit signals, but are diluted onto a planetary scale. Follow-up of these systems can be achieved with high-precision multi-color photometry to investigate the color dependence of the transit/eclipse depth.

Here, we focused on EBs and BEBs, but it can be assumed hierarchical eclipsing binaries lead to similar numbers as BEBs, as shown for example in the yield simulations for TESS by \cite{Sullivan2015}. Additionally considering ${\sim}10^2$ false alarms, we expect $\SI{97}{\percent}$ of all initially detected NGTS transit signals are caused by false positives and false alarms. 
After the candidate vetting, we expect to remain with $\SI{82}{\percent}$ of NGTS planet candidates being caused by false positives that need to be identified by follow-up.
In comparison, CoRoT's initial detections contained $\SI{98}{\percent}$ of false positives \citep{Almenara2009}. After the vetting process the follow-up candidates included $\SI{88}{\percent}$ of false positives. 
Existing ground-based surveys like WASP and HAT are even more limited in detecting false positives in their photometric data. In the RV and photometric vetting of HAT candidates a typical frequency of $\SI{95}{\percent}$ of false positives was found \citep{Latham2009,Hartman2011}.
In their estimations for TESS' full-frame images mode \citep{Sullivan2015} find a contamination by false positives of $\SI{97}{\percent}$ in the detected signals and $\SI{81}{\percent}$ after the ad-hoc vetting process. 

Dilution by background stars can lead to an underestimation of planetary radii around target stars.
Especially planets in binary systems are more difficult to detect and may appear smaller, as the light from the binary companion decreases the transit signal further. Hence, gas giants may be misclassified as Neptunes or Super Earths. Here, we do not treat these diluted planets as false negatives, but the findings raise awareness of the importance of follow-up measurements to resolve multiple star systems and blended objects.

Precise knowledge of neighboring objects as well as target star radii is crucial for the vetting and characterization of planets.
Current results from the Gaia-ESO survey \citep{Gilmore2012,Randich2013} provide astrometric information on more than two million stars, enabling the screening of NGTS targets for nearby background stars. Upcoming data releases will provide precise parallax measurements, enhancing the precision on spectroscopic properties of target stars.


\section{Conclusion}
\label{s:Conclusion}
We developed a comprehensive simulation to investigate the impact of observing strategies, target fields, and noise properties on the planet and false positive yields of transit survey programs. We considered the NGTS facility, and showed that the yield is strongly dependent on the red noise level and detection threshold.
 
According to our simulation we show NGTS will fulfill its design purpose by finding $\sim\SIrange{240}{320}{}$ close-in planets and providing a new sample of $\sim\SIrange{40}{110}{}$ characterizable Neptune-sized and smaller planets for the anticipated four-year survey.


\section*{Acknowledgements}
\addcontentsline{toc}{section}{Acknowledgements}
We thank Peter Weathley and Simon Walker for introducing us to their previous estimations of the NGTS yield and the NGTS data, as well as Richard West, James McCormac, and Daniel Bayliss for useful discussions about NGTS.
We are grateful to Ed Gillen for helpful comments on the manuscript. We further thank Courtney Dressing for providing a fine-grid version of the occurrence rate of small planets around small stars.
Figures \ref{fig:white noise}, \ref{fig:total noise}, \ref{fig:planet_fp_Rp_P}, and \ref{fig:planets_MagV_k} were created using the cubehelix color scheme provided in \cite{Green2011}.
Maximilian N. G{\"u}nther is supported by the UK Science and Technology Facilities Council (STFC) as well as the Isaac Newton Studentship.



\bibliographystyle{mnras}
\bibliography{Guenther2016_References}

\begin{thebibliography}{}
\makeatletter
\relax
\def\mn@urlcharsother{\let\do\@makeother \do\$\do\&\do\#\do\^\do\_\do\%\do\~}
\def\mn@doi{\begingroup\mn@urlcharsother \@ifnextchar [ {\mn@doi@}
  {\mn@doi@[]}}
\def\mn@doi@[#1]#2{\def\@tempa{#1}\ifx\@tempa\@empty \href
  {http://dx.doi.org/#2} {doi:#2}\else \href {http://dx.doi.org/#2} {#1}\fi
  \endgroup}
\def\mn@eprint#1#2{\mn@eprint@#1:#2::\@nil}
\def\mn@eprint@arXiv#1{\href {http://arxiv.org/abs/#1} {{\tt arXiv:#1}}}
\def\mn@eprint@dblp#1{\href {http://dblp.uni-trier.de/rec/bibtex/#1.xml}
  {dblp:#1}}
\def\mn@eprint@#1:#2:#3:#4\@nil{\def\@tempa {#1}\def\@tempb {#2}\def\@tempc
  {#3}\ifx \@tempc \@empty \let \@tempc \@tempb \let \@tempb \@tempa \fi \ifx
  \@tempb \@empty \def\@tempb {arXiv}\fi \@ifundefined
  {mn@eprint@\@tempb}{\@tempb:\@tempc}{\expandafter \expandafter \csname
  mn@eprint@\@tempb\endcsname \expandafter{\@tempc}}}

\bibitem[\protect\citeauthoryear{{Almenara} et~al.,}{{Almenara}
  et~al.}{2009}]{Almenara2009}
{Almenara} J.~M.,  et~al., 2009, \mn@doi [AAP] {10.1051/0004-6361/200911926},
  \href {http://adsabs.harvard.edu/abs/2009A%26A...506..337A} {506, 337}

\bibitem[\protect\citeauthoryear{{Baglin}, {Auvergne}, {Barge}, {Buey},
  {Catala}, {Michel}, {Weiss}  \& {COROT Team}}{{Baglin}
  et~al.}{2002}]{Baglin2002}
{Baglin} A.,  {Auvergne} M.,  {Barge} P.,  {Buey} J.-T.,  {Catala} C.,
  {Michel} E.,  {Weiss} W.,   {COROT Team} 2002, in {Battrick} B.,  {Favata}
  F.,  {Roxburgh} I.~W.,   {Galadi} D.,  eds,  ESA Special Publication Vol.
  485, Stellar Structure and Habitable Planet Finding. pp 17--24

\bibitem[\protect\citeauthoryear{{Bakos}, {L{\'a}z{\'a}r}, {Papp}, {S{\'a}ri}
  \& {Green}}{{Bakos} et~al.}{2002}]{Bakos2002}
{Bakos} G.~{\'A}.,  {L{\'a}z{\'a}r} J.,  {Papp} I.,  {S{\'a}ri} P.,   {Green}
  E.~M.,  2002, \mn@doi [PASP] {10.1086/342382}, \href
  {http://adsabs.harvard.edu/abs/2002PASP..114..974B} {114, 974}

\bibitem[\protect\citeauthoryear{{Batalha} \& {Kepler Team}}{{Batalha} \&
  {Kepler Team}}{2012}]{Batalha2012}
{Batalha} N.~M.,  {Kepler Team} 2012, in American Astronomical Society Meeting
  Abstracts \#220. p. 306.01

\bibitem[\protect\citeauthoryear{{Batalha} et~al.,}{{Batalha}
  et~al.}{2010}]{Batalha2010}
{Batalha} N.~M.,  et~al., 2010, \mn@doi [ApJL] {10.1088/2041-8205/713/2/L103},
  \href {http://adsabs.harvard.edu/abs/2010ApJ...713L.103B} {713, L103}

\bibitem[\protect\citeauthoryear{{Borucki} et~al.,}{{Borucki}
  et~al.}{2010}]{Borucki2010}
{Borucki} W.~J.,  et~al., 2010, \mn@doi [Science] {10.1126/science.1185402},
  \href {http://adsabs.harvard.edu/abs/2010Sci...327..977B} {327, 977}

\bibitem[\protect\citeauthoryear{Brown}{Brown}{2003}]{Brown2003}
Brown T.~M.,  2003, ApJL, 593, L125

\bibitem[\protect\citeauthoryear{{Brown} \& {Latham}}{{Brown} \&
  {Latham}}{2008}]{Brown2008}
{Brown} T.~M.,  {Latham} D.~W.,  2008, preprint, \href
  {http://adsabs.harvard.edu/abs/2008arXiv0812.1305B} {} (\mn@eprint {arXiv}
  {0812.1305})

\bibitem[\protect\citeauthoryear{{Brown}, {Latham}, {Everett}  \&
  {Esquerdo}}{{Brown} et~al.}{2011}]{Brown2011}
{Brown} T.~M.,  {Latham} D.~W.,  {Everett} M.~E.,   {Esquerdo} G.~A.,  2011,
  \mn@doi [\aj] {10.1088/0004-6256/142/4/112}, \href
  {http://adsabs.harvard.edu/abs/2011AJ....142..112B} {142, 112}

\bibitem[\protect\citeauthoryear{{Bryson} et~al.,}{{Bryson}
  et~al.}{2013}]{Bryson2013}
{Bryson} S.~T.,  et~al., 2013, \mn@doi [\pasp] {10.1086/671767}, \href
  {http://adsabs.harvard.edu/abs/2013PASP..125..889B} {125, 889}

\bibitem[\protect\citeauthoryear{Burke et~al.,}{Burke et~al.}{2015}]{Burke2015}
Burke C.~J.,  et~al., 2015, The Astrophysical Journal, 809, 8

\bibitem[\protect\citeauthoryear{Buser \& Kurucz}{Buser \&
  Kurucz}{1978}]{Buser1978}
Buser R.,  Kurucz R.,  1978, A\&A, 70

\bibitem[\protect\citeauthoryear{Cameron}{Cameron}{2012}]{Cameron2012}
Cameron A.~C.,  2012, Nature, 492, 48

\bibitem[\protect\citeauthoryear{{Chazelas} et~al.,}{{Chazelas}
  et~al.}{2012}]{Chazelas2012}
{Chazelas} B.,  et~al., 2012, in Society of Photo-Optical Instrumentation
  Engineers (SPIE) Conference Series. p.~0, \mn@doi{10.1117/12.925755}

\bibitem[\protect\citeauthoryear{{Claret}, {Hauschildt}  \& {Witte}}{{Claret}
  et~al.}{2012}]{Claret2012}
{Claret} A.,  {Hauschildt} P.~H.,   {Witte} S.,  2012, \mn@doi [\aap]
  {10.1051/0004-6361/201219849}, \href
  {http://cdsads.u-strasbg.fr/abs/2012A%26A...546A..14C} {546, A14}

\bibitem[\protect\citeauthoryear{{Claret}, {Hauschildt}  \& {Witte}}{{Claret}
  et~al.}{2013}]{Claret2013}
{Claret} A.,  {Hauschildt} P.~H.,   {Witte} S.,  2013, \mn@doi [\aap]
  {10.1051/0004-6361/201220942}, \href
  {http://adsabs.harvard.edu/abs/2013A%26A...552A..16C} {552, A16}

\bibitem[\protect\citeauthoryear{{Deacon} et~al.,}{{Deacon}
  et~al.}{2015}]{Deacon2015}
{Deacon} N.~R.,  et~al., 2015, preprint, \href
  {http://adsabs.harvard.edu/abs/2015arXiv150904712D} {} (\mn@eprint {arXiv}
  {1509.04712})

\bibitem[\protect\citeauthoryear{Dravins}{Dravins}{1998}]{Dravins1998}
Dravins D. L.~L. M. E. Y. A.~T.,  1998, PASP, 110, 610

\bibitem[\protect\citeauthoryear{{Dressing} \& {Charbonneau}}{{Dressing} \&
  {Charbonneau}}{2015}]{Dressing2015}
{Dressing} C.~D.,  {Charbonneau} D.,  2015, \mn@doi [\apj]
  {10.1088/0004-637X/807/1/45}, \href
  {http://adsabs.harvard.edu/abs/2015ApJ...807...45D} {807, 45}

\bibitem[\protect\citeauthoryear{{Duch{\^e}ne} \& {Kraus}}{{Duch{\^e}ne} \&
  {Kraus}}{2013}]{Duchene2013}
{Duch{\^e}ne} G.,  {Kraus} A.,  2013, \mn@doi [\araa]
  {10.1146/annurev-astro-081710-102602}, \href
  {http://adsabs.harvard.edu/abs/2013ARA%26A..51..269D} {51, 269}

\bibitem[\protect\citeauthoryear{{Duquennoy} \& {Mayor}}{{Duquennoy} \&
  {Mayor}}{1991}]{Duquennoy1991}
{Duquennoy} A.,  {Mayor} M.,  1991, \aap, \href
  {http://adsabs.harvard.edu/abs/1991A%26A...248..485D} {248, 485}

\bibitem[\protect\citeauthoryear{Fortney, Baraffe  \& Militzer}{Fortney
  et~al.}{2011}]{Fortney2011}
Fortney J.,  Baraffe I.,   Militzer B.,  2011, Exoplanets.
University of Arizona Press

\bibitem[\protect\citeauthoryear{{Fressin} et~al.,}{{Fressin}
  et~al.}{2013}]{Fressin2013}
{Fressin} F.,  et~al., 2013, \mn@doi [ApJ] {10.1088/0004-637X/766/2/81}, \href
  {http://adsabs.harvard.edu/abs/2013ApJ...766...81F} {766, 81}

\bibitem[\protect\citeauthoryear{{Gilliland} et~al.,}{{Gilliland}
  et~al.}{2011}]{Gilliland2011}
{Gilliland} R.~L.,  et~al., 2011, \mn@doi [ApJS] {10.1088/0067-0049/197/1/6},
  \href {http://adsabs.harvard.edu/abs/2011ApJS..197....6G} {197, 6}

\bibitem[\protect\citeauthoryear{{Gilmore} et~al.,}{{Gilmore}
  et~al.}{2012}]{Gilmore2012}
{Gilmore} G.,  et~al., 2012, The Messenger, \href
  {http://adsabs.harvard.edu/abs/2012Msngr.147...25G} {147, 25}

\bibitem[\protect\citeauthoryear{{Girardi}, {Groenewegen}, {Hatziminaoglou}  \&
  {da Costa}}{{Girardi} et~al.}{2005}]{Girardi2005}
{Girardi} L.,  {Groenewegen} M.~A.~T.,  {Hatziminaoglou} E.,   {da Costa} L.,
  2005, \mn@doi [AAP] {10.1051/0004-6361:20042352}, \href
  {http://cdsads.u-strasbg.fr/abs/2005A%26A...436..895G} {436, 895}

\bibitem[\protect\citeauthoryear{{Green}}{{Green}}{2011}]{Green2011}
{Green} D.~A.,  2011, Bulletin of the Astronomical Society of India, \href
  {http://adsabs.harvard.edu/abs/2011BASI...39..289G} {39, 289}

\bibitem[\protect\citeauthoryear{{Grether} \& {Lineweaver}}{{Grether} \&
  {Lineweaver}}{2006}]{Grether2006}
{Grether} D.,  {Lineweaver} C.~H.,  2006, \mn@doi [\apj] {10.1086/500161},
  \href {http://adsabs.harvard.edu/abs/2006ApJ...640.1051G} {640, 1051}

\bibitem[\protect\citeauthoryear{{Hartman}, {Bakos}  \& {Torres}}{{Hartman}
  et~al.}{2011}]{Hartman2011}
{Hartman} J.~D.,  {Bakos} G.~{\'A}.,   {Torres} G.,  2011, in European Physical
  Journal Web of Conferences. p.~2002 (\mn@eprint {arXiv} {1011.5659}),
  \mn@doi{10.1051/epjconf/20101102002}

\bibitem[\protect\citeauthoryear{{Holman} \& {Wiegert}}{{Holman} \&
  {Wiegert}}{1999}]{Holman1999}
{Holman} M.~J.,  {Wiegert} P.~A.,  1999, \mn@doi [\aj] {10.1086/300695}, \href
  {http://adsabs.harvard.edu/abs/1999AJ....117..621H} {117, 621}

\bibitem[\protect\citeauthoryear{Koch et~al.,}{Koch et~al.}{2010}]{Koch2010}
Koch D.~G.,  et~al., 2010, The Astrophysical Journal Letters, 713, L79

\bibitem[\protect\citeauthoryear{Kurucz}{Kurucz}{1993}]{Kurucz1993}
Kurucz R.,  1993

\bibitem[\protect\citeauthoryear{Latham et~al.,}{Latham
  et~al.}{2009}]{Latham2009}
Latham D.~W.,  et~al., 2009, The Astrophysical Journal, 704, 1107

\bibitem[\protect\citeauthoryear{{Lendl}, {Gillon}, {Queloz}, {Alonso},
  {Fumel}, {Jehin}  \& {Naef}}{{Lendl} et~al.}{2013}]{Lendl2013}
{Lendl} M.,  {Gillon} M.,  {Queloz} D.,  {Alonso} R.,  {Fumel} A.,  {Jehin} E.,
    {Naef} D.,  2013, \mn@doi [\aap] {10.1051/0004-6361/201220924}, \href
  {http://adsabs.harvard.edu/abs/2013A%26A...552A...2L} {552, A2}

\bibitem[\protect\citeauthoryear{{Lucy}}{{Lucy}}{1967}]{Lucy1967}
{Lucy} L.~B.,  1967, \zap, \href
  {http://adsabs.harvard.edu/abs/1967ZA.....65...89L} {65, 89}

\bibitem[\protect\citeauthoryear{{Mayor} et~al.,}{{Mayor}
  et~al.}{2003}]{Mayor2003}
{Mayor} M.,  et~al., 2003, The Messenger, \href
  {http://adsabs.harvard.edu/abs/2003Msngr.114...20M} {114, 20}

\bibitem[\protect\citeauthoryear{{Mazeh}}{{Mazeh}}{2008}]{Mazeh2008}
{Mazeh} T.,  2008, in {Goupil} M.-J.,  {Zahn} J.-P.,  eds,  EAS Publications
  Series Vol. 29, EAS Publications Series. pp 1--65 (\mn@eprint {arXiv}
  {0801.0134}), \mn@doi{10.1051/eas:0829001}

\bibitem[\protect\citeauthoryear{{Mu{\~n}oz} \& {Lai}}{{Mu{\~n}oz} \&
  {Lai}}{2015}]{Munoz2015}
{Mu{\~n}oz} D.~J.,  {Lai} D.,  2015, preprint, \href
  {http://adsabs.harvard.edu/abs/2015arXiv150505514M} {} (\mn@eprint {arXiv}
  {1505.05514})

\bibitem[\protect\citeauthoryear{{Norton} et~al.,}{{Norton}
  et~al.}{2011}]{Norton2011}
{Norton} A.~J.,  et~al., 2011, \mn@doi [\aap] {10.1051/0004-6361/201116448},
  \href {http://adsabs.harvard.edu/abs/2011A%26A...528A..90N} {528, A90}

\bibitem[\protect\citeauthoryear{{Pepe} et~al.,}{{Pepe}
  et~al.}{2014}]{Pepe2014}
{Pepe} F.,  et~al., 2014, preprint, \href
  {http://adsabs.harvard.edu/abs/2014arXiv1401.5918P} {} (\mn@eprint {arXiv}
  {1401.5918})

\bibitem[\protect\citeauthoryear{{Pickles}}{{Pickles}}{1998}]{Pickles1998}
{Pickles} A.~J.,  1998, \mn@doi [PASP] {10.1086/316197}, \href
  {http://cdsads.u-strasbg.fr/abs/1998PASP..110..863P} {110, 863}

\bibitem[\protect\citeauthoryear{{Pollacco} et~al.,}{{Pollacco}
  et~al.}{2006}]{Pollacco2006}
{Pollacco} D.~L.,  et~al., 2006, \mn@doi [PASP] {10.1086/508556}, \href
  {http://esoads.eso.org/abs/2006PASP..118.1407P} {118, 1407}

\bibitem[\protect\citeauthoryear{{Pont}, {Zucker}  \& {Queloz}}{{Pont}
  et~al.}{2006}]{Pont2006}
{Pont} F.,  {Zucker} S.,   {Queloz} D.,  2006, \mn@doi [MNRAS]
  {10.1111/j.1365-2966.2006.11012.x}, \href
  {http://adsabs.harvard.edu/abs/2006MNRAS.373..231P} {373, 231}

\bibitem[\protect\citeauthoryear{{Queloz} et~al.,}{{Queloz}
  et~al.}{2000}]{Queloz2000}
{Queloz} D.,  et~al., 2000, \aap, \href
  {http://adsabs.harvard.edu/abs/2000A%26A...354...99Q} {354, 99}

\bibitem[\protect\citeauthoryear{{Raghavan} et~al.,}{{Raghavan}
  et~al.}{2010}]{Raghavan2010}
{Raghavan} D.,  et~al., 2010, \mn@doi [ApJ] {10.1088/0067-0049/190/1/1}, \href
  {http://adsabs.harvard.edu/abs/2010ApJS..190....1R} {190, 1}

\bibitem[\protect\citeauthoryear{{Randich}, {Gilmore}  \& {Gaia-ESO
  Consortium}}{{Randich} et~al.}{2013}]{Randich2013}
{Randich} S.,  {Gilmore} G.,   {Gaia-ESO Consortium} 2013, The Messenger, \href
  {http://adsabs.harvard.edu/abs/2013Msngr.154...47R} {154, 47}

\bibitem[\protect\citeauthoryear{Slawson et~al.,}{Slawson
  et~al.}{2011}]{Slawson2011}
Slawson R.~W.,  et~al., 2011, The Astronomical Journal, 142, 160

\bibitem[\protect\citeauthoryear{{Soszy{\'n}ski} et~al.,}{{Soszy{\'n}ski}
  et~al.}{2015}]{Soszynski2015}
{Soszy{\'n}ski} I.,  et~al., 2015, \actaa, \href
  {http://adsabs.harvard.edu/abs/2015AcA....65...39S} {65, 39}

\bibitem[\protect\citeauthoryear{{Sullivan} et~al.,}{{Sullivan}
  et~al.}{2015}]{Sullivan2015}
{Sullivan} P.~W.,  et~al., 2015, preprint, \href
  {http://adsabs.harvard.edu/abs/2015arXiv150603845S} {} (\mn@eprint {arXiv}
  {1506.03845})

\bibitem[\protect\citeauthoryear{Walker}{Walker}{2013}]{Walker2013}
Walker S.,  2013, PhD thesis, University of Warwick

\bibitem[\protect\citeauthoryear{{Weiss} \& {Marcy}}{{Weiss} \&
  {Marcy}}{2014}]{Weiss2014}
{Weiss} L.~M.,  {Marcy} G.~W.,  2014, \mn@doi [ApJL]
  {10.1088/2041-8205/783/1/L6}, \href
  {http://adsabs.harvard.edu/abs/2014ApJ...783L...6W} {783, L6}

\bibitem[\protect\citeauthoryear{{Wheatley} et~al.,}{{Wheatley}
  et~al.}{2013}]{Wheatley2013}
{Wheatley} P.~J.,  et~al., 2013, in European Physical Journal Web of
  Conferences. p. 13002 (\mn@eprint {arXiv} {1302.6592}),
  \mn@doi{10.1051/epjconf/20134713002}

\bibitem[\protect\citeauthoryear{Winn}{Winn}{2011}]{Winn2011}
Winn J.~N.,  2011, Exoplanets.
University of Arizona Press

\makeatother
\end{thebibliography}

\clearpage
\appendix

\section{Input file for the yield simulation on the example of NGTS}
\label{app:Input file for the yield simulation on the example of NGTS}
\begin{footnotesize}
\begin{Verbatim}
######################################################################
# Target list criteria
######################################################################
15.		#MagV_target = 15.			#upper limit on MagV for target stars
22.5		#magV_limit = 22.5			#upper limit on MagV for background stars
2.		#Rs_limit = 2.				#upper limit on stellar radius of target stars (in Rsun)
10000.		#Teff_limit = 10000.			#upper limit on effective temperature of target stars (in K)
6.5		#logg_limit = 6.5			#upper limit on logg of target stars
######################################################################
# CCD resolution
######################################################################
50331648.	#pixel_number = 12. * 2048.**2		#number of pixels in total; 12 cameras with 2048^2 pixel
4194304.	#pixel_per_ccd = 2048*2048		#number of pixels per CCD; 1 camera with 2048^2 pixel
######################################################################
# Observing strategy
######################################################################
88.8        	#FoV = 12. * 7.4.			#FoV that is observed (in sq.deg.)
122.	        #FoV_duration = 365./3.			#time spent per FoV (in days)
12.	        #N_FoVs = 4.*3.				#number of FoVs surveyed in total (4 years * 3 fields per year)
7.	        #hours_per_night = 7.			#average number of hours observed per night 
######################################################################
# Instrument parameters, noise, aperture
######################################################################
10. 	        #exposure = 10.				#exposure (in s)
1.49		#readout = 1.49				#CCD readout time (in s) 
0.2		#D_tel = 0.2				#telescope aperture (in m)
1.5        	#airmass = 1.5				#average airmass
2400.		#h_tel = 2400.				#height of the telescope above sea level (in m)
125.        	#noise_sky = 125.			#sky noise (in e- per s per pixel)
0.06       	#noise_dark = 0.06			#dark noise (in e- per s per pixel)
10.		#noise_readout = 10.			#readout noise (in e- per pixel)
28.2743338823	#N_apert_pixel = np.pi*(3.**2.)		#number of pixels in aperture (in average)
######################################################################
# Detection criteria
######################################################################
5.		#detection_threshold = 5.		#detection threshold DT (planet detected if signal-to-noise > DT)
0.001		#red_noise = 1./1000. 			#as fraction (1/1000 ~ 1 mmag)
3.		#num_transits_threshold = 3.		#minimum number of visible transits required for detection
\end{Verbatim}
\end{footnotesize}




\bsp	
\label{lastpage}
\end{document}